\documentclass{elsart}

\usepackage{graphicx}

\usepackage{amssymb}

\begin{document}

\begin{frontmatter}



\title{State Transitions and the Continuum Limit 
for a 2D Interacting, Self-Propelled Particle System}


\author[duke,ucla]{Yao-li Chuang\corauthref{cor1}}\ead{chuang@phy.duke.edu, chuang@math.ucla.edu},
\author[ucla]{Maria R. D'Orsogna}\ead{dorsogna@math.ucla.edu},
\author[northorp]{Daniel Marthaler}\ead{daniel.marthaler@ngc.com},
\author[duke,ucla]{Andrea L. Bertozzi}\ead{bertozzi@math.ucla.edu}, and
\author[ucla]{Lincoln S. Chayes}
\address[duke]{Department of Physics, Duke University, Durham, NC, USA}
\address[ucla]{Department of Mathematics, UCLA, Los Angeles, CA, USA}
\address[northorp]{ACS-UMS, Northrop Grumman Corp, Rancho Bernardo, CA, USA}
\corauth[cor1]{Corresponding author}

\begin{abstract}
We study a class of {\em swarming problems} wherein particles evolve 
dynamically via pairwise
interaction potentials and a velocity selection mechanism. 
We find that the swarming system undergoes various changes of state as
a function of the self-propulsion and interaction potential parameters.
In this paper, we utilize a procedure which, in a definitive way, 
connects a class of individual-based models to their continuum formulations 
and determine criteria for the validity of the latter.
H-stability of the interaction potential plays a fundamental role in 
determining both the validity of the continuum approximation and 
the nature of the aggregation state transitions.
We perform a linear stability analysis of the continuum model 
and compare the results to the simulations of the individual-based one.
\end{abstract}

\end{frontmatter}

\section{Introduction}
The collective behaviors of aggregating organisms are of interest in 
various fields, including biology, engineering, mathematics, and physics 
\cite{Vicsek-phase,TonerTuSwarming,Mogilner96,Sugawara97,ScienceSwarming,NaomiUAV,FishSchoolSwarming,UAV1,UAV2}. 
There are primarily two classes of pertinent models:
individual-based and continuum ones. 
In the first case, one considers a collection of $N$ individual 
entities, so that the system is defined on the ``microscopic'' scale.
Such models are particularly useful for the study 
and algorithmic design of small-size aggregates such as
 artificial swarms of autonomous vehicles.  
Larger discrete systems are adaptive to statistical analysis
\cite{Vicsek-phase,Niwa94,Shimoyama,Romey1,Brownian1,FlierlDiscreteModel,LevineModel,CouzinDiscreteModel,Brownian2,Brownian3,Viscido1,Viscido2,Brownian4}. 
Continuum models typically describe swarms through a density function 
 $\rho \left( \vec{r} \right) $
and a velocity vector field $\vec u\left( \vec{r} \right)$. 
These obey appropriate non-linear and often non-local equations. 
One may presume these equations are derived from,
or at least connected back to, the original microscopic system.  
Continuum models are useful for theoretical analysis of swarming systems   
\cite{TonerTuSwarming,Mogilner96,FlierlDiscreteModel,LevineModel,BandSolutionSwarming,Mogilner1,Oien,TopazSwarming,Topaz2}.
Although both individual-based and continuum models are applicable, 
the connection between the two has, in fact, 
not been particularly well established. 
A primary purpose of this paper is to better unify the two approaches, 
for a particular class of models, 
following the classical statistical studies of fluids 
\cite{IrvingKirkwood}.
We investigate the validity of the continuum
model by a detailed comparison 
with the associated individual based one.
In particular, for certain interaction forms, the two descriptions yield the
same morphological patterns.
Furthermore, and perhaps more importantly, we are able to explain why the 
continuum model fails qualitatively for the other cases,
where discrepancies exist.

In Ref. \cite{dorsogna}, a criterion from classical statistical mechanics 
known as {\it H-stability} was applied to individual based swarming models.  
A system of $N$ interacting particles is said to be H-stable if
 the potential energy per particle 
is bounded below by a constant 
which is independent of the number of particles present \cite{Ruelle}.  
H-stability is a necessary and sufficient condition for the existence 
of thermodynamics.
Indeed, a system without this stability will, in the thermodynamic limit, 
collapse onto itself; such systems are called {\it catastrophic}.
In Ref. \cite{dorsogna}, numerical simulations strongly suggest that 
a specific non-Hamiltonian swarming system exhibits the same  
stability trends observed in classical Hamiltonian many-body systems. 
In this paper, we show that H-stability also plays an
important role in determining the correct passage to
the continuum limit.

This paper is organized as follows. 
In Sec.\,\ref{sec:IBM}, the individual-based model is presented.
We study various aggregation states and transitions between them 
via numerical simulations.
In Sec.\,\ref{sec:Continuum-model}, a continuum model is derived.
In Sec.\,\ref{sec:comparison},
we quantitatively compare steady states
of both continuum and discrete models.  We show that,
 while the proposed continuum model works well 
in the catastrophic regime, discrepancies arise for large H-stable systems.
In Sec.\,\ref{sec:linear-stability}, the stability of the homogeneous solution
of the continuum model is studied and compared to the numerical results of 
the individual-based model.
In Sec.\,\ref{Sec-Discussion}, we discuss the choices between soft-core and
hard-core interaction potentials.

\section{The individual-based model \label{sec:IBM}}
%
\subsection*{Background}

\begin{figure}
\includegraphics[%
  width=1.0\columnwidth]{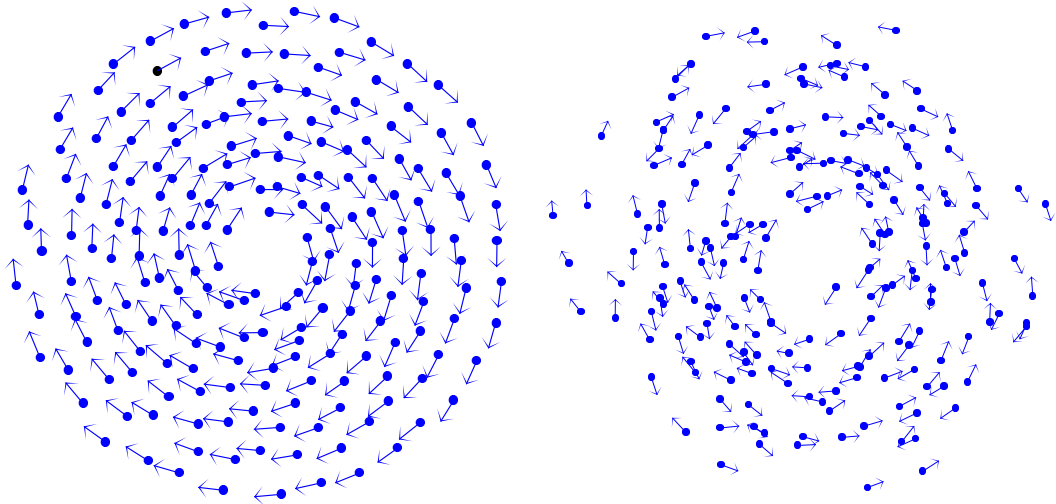}

\caption{\label{cap:single-double}Left: The swarming pattern of a single
mill. Right: The swarming pattern of two interlocking mills.}
\end{figure}

Common swarming patterns have been observed and reported in
various species in nature. 
One example is a coherent flock formation involving a polarized group 
moving in the same direction. 
Another example is a single rotating mill pattern, 
with a rather stationary center of mass, as in
Fig.\,\ref{cap:single-double} (left).
The rotating-mill pattern is frequently observed in both 
two and three dimensions among many species and across different sizes 
\cite{ScienceSwarming,FishSchoolSwarming,Schneirla}.
Various individual-based models have been able to reproduce these patterns 
within certain parameter ranges 
\cite{Brownian1,LevineModel,CouzinDiscreteModel,Brownian2,Brownian3,Brownian4}.
An unusual pattern of overlapping double mills is also reported 
in Ref. \cite{LevineModel}, similar to the simulation shown in 
Fig.\,\ref{cap:single-double} (right). 
The double-mill phenomenon is observed in the early
stages of aggregation of {\em Myxococcus
xanthus\/}, a single-cell bacteria driven by self-propelling motors
\cite{Koch-White-myxobacteria}. 
We show that this configuration can be obtained using the same swarming 
mechanism that produces the single-mill pattern 
but exists in a different parameter regime.
The rarity of the double-mill state is discussed in Sec. \ref{Sec-Discussion}.

\subsection*{Equations of motion}

The swarming model we present in this paper is described by the following
equations of motion
\begin{eqnarray}
\frac{\mathrm{d}\vec{x}_{i}}{\mathrm{d}t} & = & 
\vec{v}_{i}\;, \label{eq:OurPosEq}\\
m_{i}\frac{\mathrm{d}\vec{v}_{i}}{\mathrm{d}t} & = & 
\alpha \vec{v}_{i} - \beta \left| \vec{v}_{i} \right|^{2} \vec{v}_{i}
-\vec{\nabla} U_{i}, \label{eq:OurMomEq}
\end{eqnarray}
where $m_{i}$, $\vec{x}_{i}$ and $\vec{v}_{i}$ are, 
respectively, the mass, position,
and velocity of particle $i$. 
The terms $\alpha\vec{v}_{i}$ and 
$-\beta\left|\vec{v}_{i}\right|^{2}\vec{v}_{i}$ define the mechanism 
of self-acceleration and deceleration which give the particles a tendency 
to approach an equilibrium speed $v_\mathrm{eq} = \sqrt{{\alpha/\beta}}$. 
This Rayleigh-type dissipation was originally proposed in Ref. \cite{Rayleigh} 
and is often used in the literature as a velocity-selecting mechanism 
\cite{TonerTuSwarming,Niwa94,Brownian1,Brownian2,Brownian3,Brownian4,Weihs73}.
The potential $U_{i}$ describes the interaction of particle $i$ with the
other particles.
One common choice is the following 
\cite{LevineModel,Brownian3,Mogilner1,dorsogna}
\begin{eqnarray}
U_{i} \equiv U \left( \vec{x}_i \right) 
& = & \sum_{j\ne i} { V \left( \left| \vec{x}_i-\vec{x}_j \right| \right) } 
 = \sum_{j\ne i}
{ \left( -C_a \e^{-\frac{\left| \vec{x}_i-\vec{x}_j \right|}{\ell_a}}
+ C_r \e^{-\frac{\left| \vec{x}_i-\vec{x}_j \right|}{\ell_r}} \right)}.
\label{eq:interaction_term}
\end{eqnarray}
Eq.\,(\ref{eq:interaction_term}) assumes that only pairwise
interactions are significant and ignores $N$-body
interactions with $N\ge3$. 
The pairwise interaction consists of an attraction and a repulsion 
with $C_a$, $C_r$ specifying their respective strengths and 
$\ell_a$, $\ell_r$ their effective interaction length scales. 
Similar behaviors are also observed with other functional forms of
interaction potential that are characteristically similar to 
Eq.\,(\ref{eq:interaction_term}).
Note that to simplify the analysis, our model is deterministic. 
Stochastic forces appear in many other models 
\cite{Vicsek-phase,Niwa94,Brownian1,Brownian2,Brownian3,Brownian4}.
In our simulations, we observe
that noise affects the swarming patterns only beyond certain thresholds,
and thus its consequences are not investigated.

\begin{figure}
\includegraphics[%
  width=1.0\columnwidth]{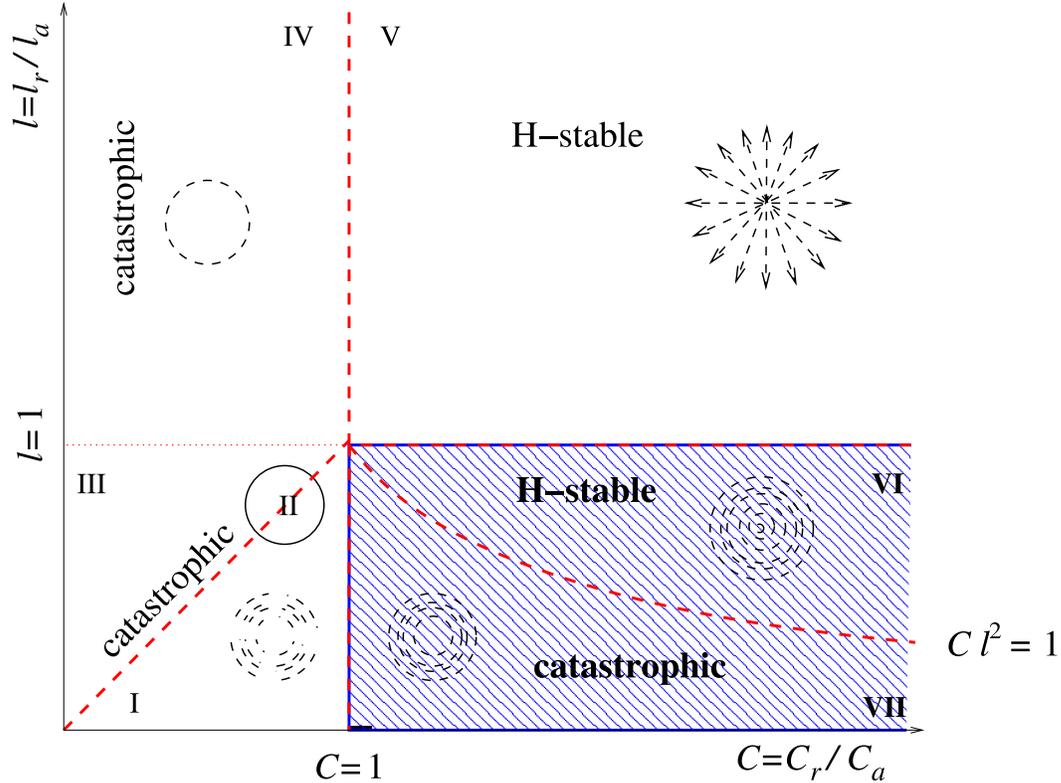}

\caption{\label{cap:H-stab-diag}The H-stability diagram of the interaction
potential in Eq.\,(\ref{eq:interaction_term}) \cite{dorsogna}. 
The shaded region is the so-called biologically relevant region where the 
interaction consists of a long-range attraction and a short-range repulsion.}
\end{figure}

We can non-dimensionalize the equations of motion by substituting 
$t^{\prime} = \left( m_i / {\ell_a}^2 \beta \right) t$, 
${\vec{x}_i}^{\prime} = \vec{x}_i / \ell_a$,
and thus, ${\vec{v}_i}^\prime = \left( \ell_a \beta / m_i \right) \vec{v}_i$ 
into Eqs.\,(\ref{eq:OurPosEq})\,-\,(\ref{eq:interaction_term})
\begin{eqnarray}
\frac{\mathrm{d}{\vec{x}_{i}}^\prime}{\mathrm{d}t^\prime} & = & 
{\vec{v}_{i}}^\prime\;, \label{eq:nondimPosEq} \\
\frac{\mathrm{d}{\vec{v}_{i}}^\prime}{\mathrm{d}t^\prime}
& = & 
\alpha^\prime {\vec{v}_{i}}^\prime 
- \left| {\vec{v}_{i}}^\prime \right|^{2} {\vec{v}_{i}}^\prime
- \frac{1}{{m_i}^\prime} \vec{\nabla}_{{\vec{x}_i}^\prime} {U_{i}}^\prime , 
\label{eq:nondimMomEq}\\
{U_{i}}^\prime  
& = & \sum_{j\ne i}
{ \left( - \e^{-\left| {\vec{x}_i}^\prime - {\vec{x}_j}^\prime \right|}
+ C \e^{-\frac{\left| {\vec{x}_i}^\prime - {\vec{x}_j}^\prime \right|}{\ell}} 
\right)}, 
\label{eq:nondiminteraction}
\end{eqnarray}
where $\alpha^\prime = \left( \alpha \beta {\ell_a}^2 \right) / {m_i}^2 $, 
${m_i}^\prime = {m_i}^3 / \left( \beta^2 C_a {\ell_a}^2 \right) $, 
$C = C_r / C_a$, and $\ell = \ell_r / \ell_a$;
hence, the model is essentially a 4-parameter one.
In Ref. \cite{dorsogna} the 
effects of varying $C$ and $\ell$, 
which affect H-stability, are explored.
In this paper we investigate the role of $\alpha^\prime$, 
the relative strength of the self-driving force with respect to the 
interaction.
The parameter ${m_i}^\prime$ affects the time scale of the particle
interaction and is fixed during our investigations.
Note that the dimensional parameter $\alpha$ only appears in the dimensionless
parameter $\alpha^\prime$, 
which allows us to vary $\alpha$ to change $\alpha^\prime$  
without affecting the other three independent parameters, 
provided that $\beta$, $\ell_a$, and $m_i$ are fixed during the process.
To preserve the original meaning of the model parameters, 
our results are presented in the dimensional form by using 
Eqs.\,(\ref{eq:OurPosEq})\,-\,(\ref{eq:interaction_term}). 
Only the {\em biologically relevant} cases that consist of 
a long-range attraction and a short-range repulsion are studied. 
In other words, we confine our analysis to the parameter space 
where $C>1$ and $\ell<1$, 
which is shown by the shaded region of the H-stability phase diagram in 
Fig.\,\ref{cap:H-stab-diag}. 
The extremely collapsing cases reported in \cite{dorsogna}, 
such as the ring formations and the clump formations illustrated in other
regions, do not change morphology with respect to $\alpha^\prime$.

\subsection*{Swarming states}

\begin{figure}
\includegraphics[%
  width=1.0\columnwidth]{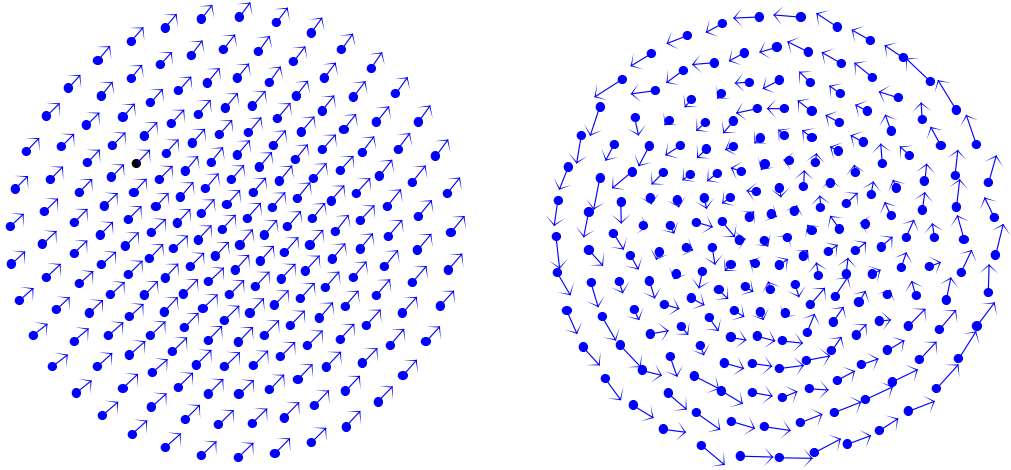}

\caption{\label{cap:hstable-mill} Left: The coherent flock state. Right: The
rigid-body rotation state.}
\end{figure}

We use the fourth order Runge-Kutta and the four step Adam-Bashforth
methods for the numerical simulation of 
Eqs.\,(\ref{eq:OurPosEq})\,-\,(\ref{eq:interaction_term}) 
\cite{Lambert:num-ode}.
We impose free boundary conditions to the model and 
initiate the simulation with random distributions of particle
position and velocity.
Figure\,\ref{cap:single-double} shows two typical patterns 
akin to those observed in various natural swarms. 
On the left panel is the {\em single-mill state\/}, 
where every particle travels at the same speed $v_\mathrm{eq}$ around an 
empty core at the center of the swarm.
On the right panel is the {\em double-mill state\/}, 
in which particles travel in both clockwise and counterclockwise directions,
also at a uniform speed $v_\mathrm{eq}$. 
In this second example, when viewed as two superimposed mills,
the cores of each mill do not exactly coincide 
but rather fluctuate near each other.
Another two states are shown in Fig.\,\ref{cap:hstable-mill}.
On the left panel is the {\em coherent flock state\/}.
All particles travel at a unified velocity 
(i.e., with the same speed and direction) 
while self-organizing into a stable formation.
On the right panel is the {\em rigid-body rotation state\/}.
The flock formation closely resembles that of the coherent flock,
but instead of traveling at the same velocity, 
the particles circulate around
the swarm center defining a constant angular velocity $\omega$.
Unlike the single and double-mill state, 
where particles swim freely within the swarm, 
both the coherent flock and the rigid-body rotation states bind particles
at fixed relative positions, exhibiting a lattice-type formation.
Hence, we also use the term {\em lattice states\/} to refer to both
the coherent flock and the rigid-body rotation states.
Note that the coherent flock is a traveling wave
solution of the model, and thus
 a solution of the following Euler-Lagrange equation
\begin{displaymath}
\vec{\nabla} U_i  =  \vec{\nabla} \sum_{j\ne i}
{ V \left( \left| \vec{x}_i-\vec{x}_j \right| \right) } = 0.
\end{displaymath}
It is interesting to note that this equation 
arises in the context of a gradient flow algorithm for autonomous 
vehicle control \cite{UAV1,passino1,passino2,passino3}. 
Thus the flock formations have the shape and structure as equilibria
of the gradient flow problem with the same potential.

The coherent flock and the single-mill states are among the most 
common patterns observed in biological swarms 
\cite{ScienceSwarming,FishSchoolSwarming,Schneirla}.
The double-mill pattern is also occasionally seen; 
an example is the {\em M. xanthus\/} bacteria at the onset of fruiting
body formation \cite{Koch-White-myxobacteria}.
On the other hand, natural occurrences of rigid-body rotation, to the best
of our knowledge, have not been reported in the literature. 
Indeed, the rigid-body rotation, 
where every particle travels at a constant angular velocity $\omega$, 
does not define a rotationally symmetric solution for 
Eqs.\,(\ref{eq:OurPosEq})\,-\,(\ref{eq:interaction_term})
and the swarm is observed to drift randomly due to the unbalanced 
self-driving mechanism.
The random drift may eventually break the rotational symmetry and 
turn the swarm into a coherent flock after a transient period.
Thus, we speculate that this pattern may only be a meta-stable 
or a transient state.
In addition to the above aggregation states, 
the particles may simply escape from the collective potential field,
and no aggregation is observed. 
We name it the {\em dispersed state\/}.

Using numerical simulations, we find that the H-stable swarms undergo
a different state transition process from that of the non-H-stable 
swarms.
For both H-stable and catastrophic interactions, the lattice states,
as shown in Fig.\,\ref{cap:hstable-mill}, 
emerge for low values of $\alpha$, and thus, of low $v_\mathrm{eq}$. 
In this case, the confining interaction potential is stronger than the 
kinetic energy of individual particles and tends to bind the particles 
at specific ``crystal'' lattice sites.
Most initial conditions lead to the coherent flock state 
while some occasions result in the rigid-body rotation state. 
The state transition of H-stable swarms is simpler. 
As $\alpha$ increases, the particles eventually gain enough kinetic energy
to dissolve the aggregation.

\begin{figure}
\includegraphics[%
  width=1.0\columnwidth]{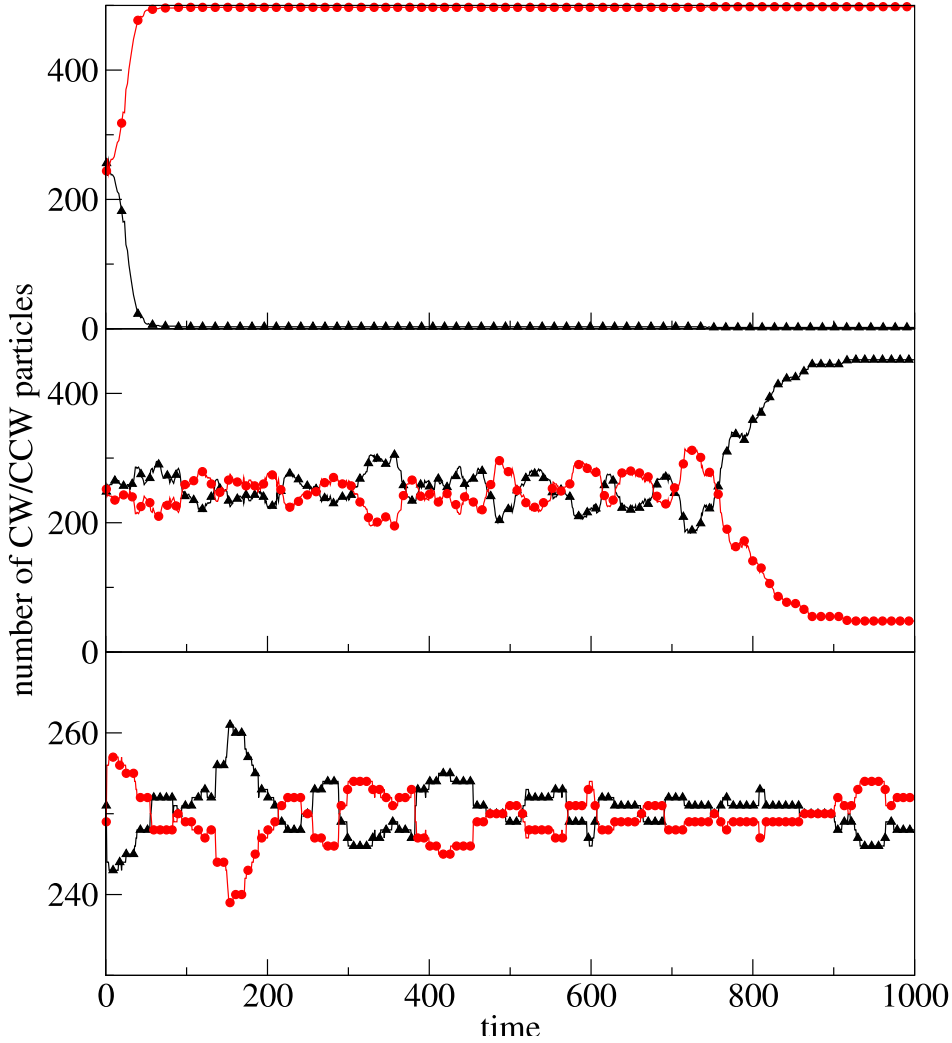}

\caption{\label{cap:counting} Time variation of the numbers of particles
rotating in different directions: The triangles represent the number of 
CCW particles while the circles are of CW ones. 
(Top) $\alpha=1.5$. (Middle) $\alpha=4.0$. (Lower) $\alpha=6.0$. 
The fixed parameters are $\beta =0.5$, $C_a=0.5$, $C_r=1.0$,
$\ell_a=2.0$, $\ell_r=0.5$, and $N=500$. 
All parameters here and throughout the paper are in arbitrary units.}
\end{figure}

The state transition of catastrophic swarms is characterized by more
behavioral stages. 
Starting from the lattice states and upon increasing $\alpha$, 
the particles gain more kinetic energy from the environment 
to reach $v_\mathrm{eq}$ 
and are able to break away from the crystal lattice sites. 
However, unlike H-stable swarms, the interaction potential in the catastrophic
regime is still strong enough to aggregate
 medium-speed particles within a swarm.
In this regime, core-free mill states emerge, 
as shown in Fig.\,\ref{cap:single-double}.
Since all particles travel at a non-zero uniform speed, 
the centripetal force provided by the collective interaction potential
is not strong enough to sustain such particles too close to 
the rotational center. 
As a result, the mill core is a particle-free region.
At moderate $\alpha$, a single mill state emerges.  At
slightly higher $\alpha$, we observe both single mills
and double mills as possible states.
In the latter case,
 the interaction potential gradually loses its effectiveness to
unify the clockwise (CW) and counterclockwise (CCW) rotational directions; 
particles traveling in the opposite direction with respect to the majority
tend to not change their direction of motion, and double mills can emerge.
The transition from single to double mill is a gradual process. 
Figure\,\ref{cap:counting} shows the number of particles 
in each rotational direction for various values of $\alpha$. 
In the single-mill regime, particles traveling at one direction are quickly 
assimilated into the other (Fig.\,\ref{cap:counting},\,top). 
Upon increasing $\alpha$, the particles no longer settle into a unified
rotational direction (Fig.\,\ref{cap:counting},\,middle),  
and for large enough $\alpha$, approximately the same number of particles 
travel in each of the CW and CCW directions 
(Fig.\,\ref{cap:counting},\,bottom). 
The presence of either a velocity alignment rule or a hard-core repulsive
interaction will destroy this double-mill state.
The latter case is 
because hard-cores always provide a system with H-stability.
Thus, it is clear that for sufficiently many particles, the double mills
will ultimately break apart.
Notwithstanding, it appears that the double mills are especially sensitive
to hard-cores and, even for small cores and moderate $N$, we have not 
observed these structures.
As for the coherent flock state, it still remains a possibility in this regime
where the mill states occur.
However, the basin of attraction is greatly reduced, and 
only very polarized initial conditions can lead to 
the coherent flock formation.
As $\alpha$ increases beyond the double-mill regime, 
particle kinetic energy eventually becomes high enough to break up the swarm.
This is the dispersed state, and no aggregation can be found.
%

\begin{figure}
\includegraphics[%
  width=1.0\columnwidth]{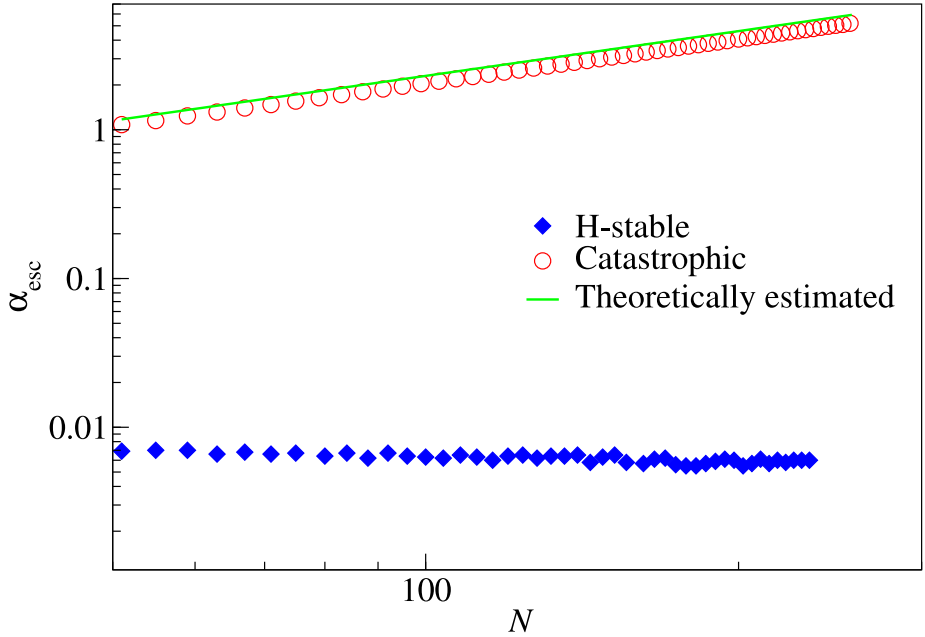}

\caption{\label{cap:escap-alpha} $\alpha_\mathrm{esc}$ versus the total
number of particles in an H-stable swarm ($\beta=0.5$, $C_a=0.5$,
$C_r=1.0$, $\ell_a=2.0$, $\ell_r=1.5$, dashed line) compared to
that of a catastrophic swarm ($\beta=0.5$, $C_a=0.5$, $C_r=1.0$, 
$\ell_a=2.0$, $\ell_r=0.5$, dotted line). The solid line is the curve estimated
by Eq.\,(\ref{eq:escaping-alpha}).}
\end{figure}

Upon fixing the other parameters, the threshold between the aggregation
and the dispersed states is described by a critical escape value of $\alpha$,
denoted by $\alpha_\mathrm{esc}$.
Figure\,\ref{cap:escap-alpha} shows $\alpha_\mathrm{esc}$ of an H-stable 
swarm versus a catastrophic one, in which 
single-mill states are generated and $\alpha$ is increased until 
the dispersed regime is attained. 
For the H-stable case, 
$\alpha_\mathrm{esc}$ does not vary significantly with respect 
to the total particle number of the swarm, denoted by $N$, 
due to the fact that the nearest neighbor distance ($\delta_\mathrm{NND}$) 
does not change as $N$ increases.
As a result, the binding potential energy of the interaction force 
acting over each particle is independent of $N$. 
On the other hand, $\alpha_\mathrm{esc}$ of the catastrophic swarm varies
linearly with respect to $N$. 
From our numerical simulations, we observe that the outer and the inner radii 
of the catastrophic swarm remain approximately fixed with respect to $N$  
while $\alpha\lesssim\alpha_\mathrm{esc}$. 
Based on this observation, we can derive a semi-empirical formula 
to estimate the value of $\alpha_\mathrm{esc}$  
by assuming that particles are uniformly distributed 
in a doughnut shape domain.
By balancing the centripetal and the interaction forces, we obtain
\begin{eqnarray}
\frac{m \alpha_\mathrm{esc}}{2 \beta} & = & 
\frac{N}{2 \pi \left( R_\mathrm{out}^{2}-R_\mathrm{in}^{2} \right) }
\int_{R_\mathrm{in}}^{R_\mathrm{out}}
V\left( \left| \vec{r} - R_\mathrm{out} \hat{x} \right| \right)
 \mathrm{d} \vec{r}, 
\label{eq:escaping-alpha}
\end{eqnarray}
where $R_\mathrm{in}$ and $R_\mathrm{out}$ denote the inner and the outer
radii of the single mill, respectively, 
and $\hat{x}$ is an arbitrary unit vector.
This estimate predicts that $\alpha_\mathrm{esc}$ should scale 
linearly with $N$, 
which is clearly illustrated in Fig.\,\ref{cap:escap-alpha}, 
where we use the numerically simulated 
$R_\mathrm{out}=5.2$ and $R_\mathrm{in}=1.2$ for a quantitative comparison.

\subsection*{State transitions of H-stable and catastrophic swarms}

In order to quantitatively determine whether the swarm is in a coherent flock
state or a single-mill state, Couzin et al. have proposed two measures
\cite{CouzinDiscreteModel}: 
the {\em polarity\/}, $P$, 
and the {\em (normalized) angular momentum\/}, $M$, defined as follows
\begin{eqnarray}
P & = & \left| \frac{\sum_{i=1}^{N}{\vec{v}_i}}
{\sum_{i=1}^{N}{\left| \vec{v}_i \right|}} \right|, \label{eq:polarity}\\
M & = & \left| \frac{\sum_{i=1}^{N}{\vec{r}_i \times \vec{v}_i}}
{\sum_{i=1}^{N}{\left| \vec{r}_i \right| \left| \vec{v}_i \right|}} \right|, 
\label{eq:ang-momen}
\end{eqnarray}
where $\vec{r}_i\equiv\vec{x}_i-\vec{x}_\mathrm{CM}$, 
and $\vec{x}_\mathrm{CM}$ is the position of the center of mass. 
A perfect coherent flock results in $P=1$ and $M=0$ 
while a perfect single-mill pattern results in $M=1$ and $P=0$. 
In order to distinguish the double-mill pattern, we propose an additional 
measure by modifying the normalized angular momentum
\begin{equation}
M_\mathrm{abs} = \left| 
\frac{\sum_{i=1}^{N}{\left|\vec{r}_i \times \vec{v}_i \right|}}
{\sum_{i=1}^{N}{\left| \vec{r}_i \right| \left| \vec{v}_i \right|}}
\right|. \label{eq:abs-angular-momen}
\end{equation}
If a double-mill pattern has perfectly equal numbers of particles going at each
direction with the centers of mass of both directions exactly overlap,
$M_\mathrm{abs}=1$ and $M=0$; both $M$ and $M_\mathrm{abs}$ equal to $1$ 
for a single mill. 
\begin{figure}
\begin{center}
\includegraphics[%
  height=1.4\linewidth]{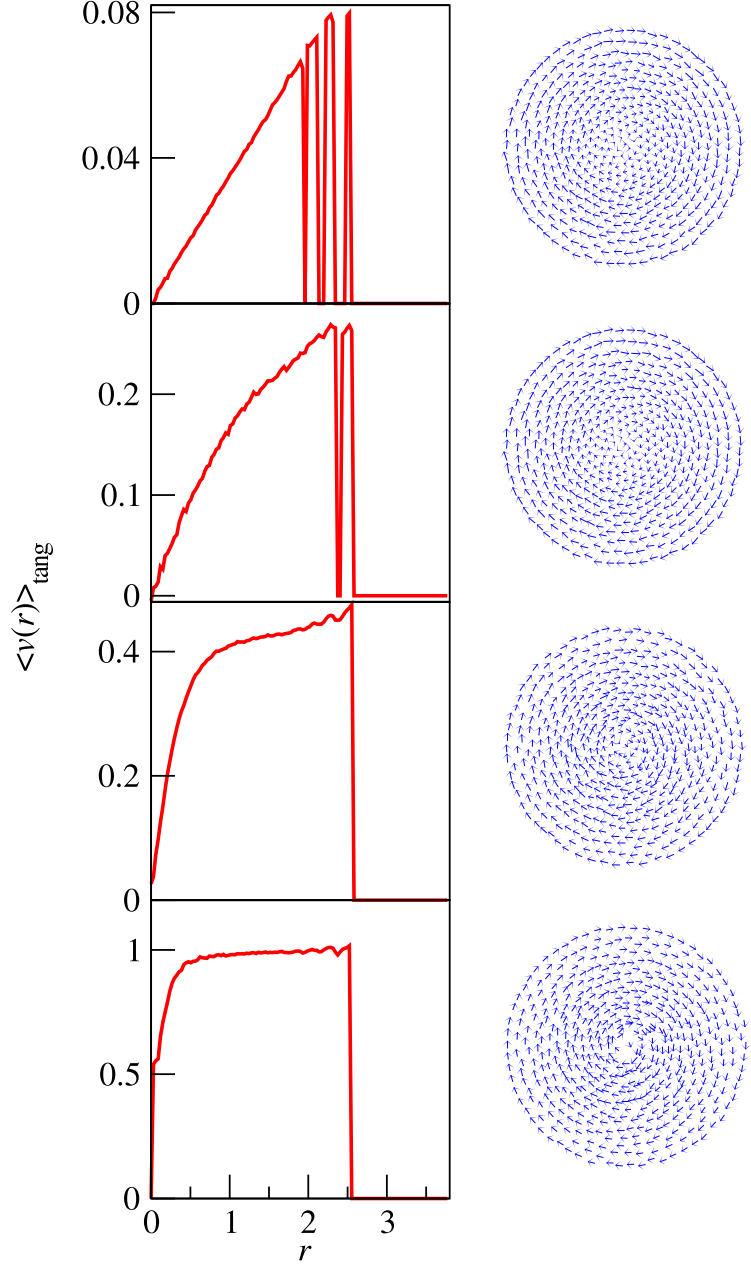}
\end{center}

\caption{\label{cap:rigid-transform} Emergence of a rotating single-mill
structure from a rigid-body rotation in the catastrophic regime. 
The left panel shows the ensemble averaged tangential velocity, 
$\langle v \left( r \right) \rangle_\mathrm{tang}$, of particles at 
a distance $r$ from the center of mass. 
Each $\langle v \left( r \right) \rangle_\mathrm{tang}$ figure corresponds
to the swarm structure of different values of $\alpha$ on the right panel:  
from top to bottom are $\alpha=0.003$, $\alpha=0.03$, $\alpha=0.1$, 
and $\alpha=0.5$. The other parameters are $\beta=0.5$, $C_a=0.5$, $C_r=1.0$, 
$\ell_a=2.0$, $\ell_r=0.5$, and $N=500$.}
\end{figure}

\begin{figure}
\includegraphics[%
  width=1.0\columnwidth]{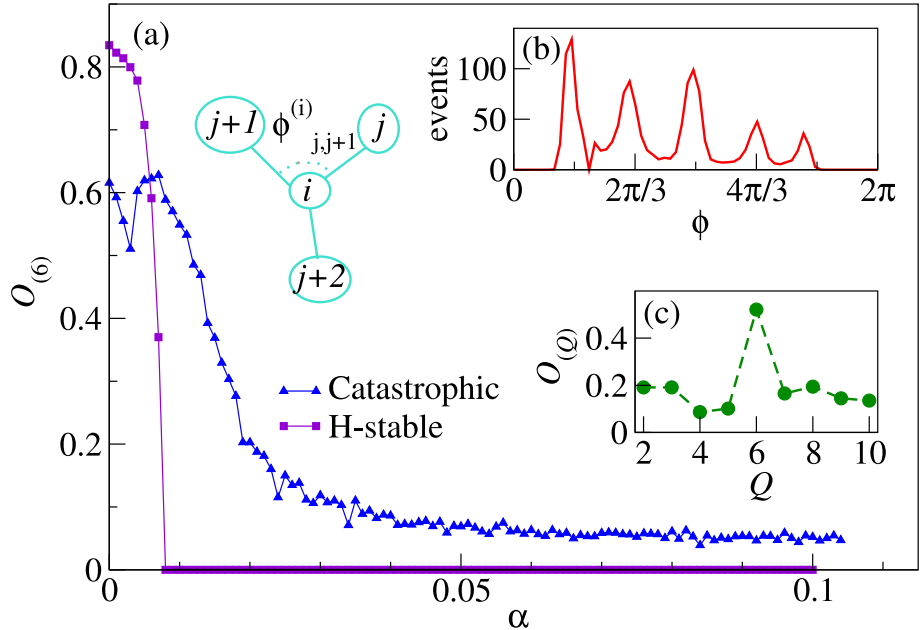}

\caption{\label{cap:ordering-factor}(a) The ordering factor of period $6$
versus $\alpha$ and an illustration showing the definition of
$\phi_{j,j+1}^{(i)}$.
The triangles are data points of the catastrophic case  
while the squares represent the H-stable case. 
The parameters other than $\alpha$ for both cases are the same as those in 
Fig.\,\ref{cap:escap-alpha} with $N=200$.
(b) The distribution of $\phi_{j,j+1}^{(i)}$ for all $i$ and $j$. 
(c) Comparison of the ordering factors of different periods $Q$. }
\end{figure}

Although the presence of the coherent flock that yields $P \simeq 1$ 
allows us to use $P$ to quantify the transition from lattice
to single-mill state, the co-existing rigid-body rotation 
state, for which $P \simeq 0$, introduces spurious events. 
Since the rigid body state has a much smaller basin of attraction than 
the coherent flock, one choice is discarding all rigid-body rotation 
events and selecting only the coherent flock ones. 
However, the boundary between a rigid body rotation and a single mill 
is ambiguous, as shown in Fig.\,\ref{cap:rigid-transform}, where 
a rigid-body rotation transforms to a single mill by increasing $\alpha$. 
Since a constant tangential speed indicates a milling formation,
and a constant angular velocity (i.e., a linear tangential speed against
$r$) characterizes a rigid-body rotation, we can see from the figure
that two states are mixed during the transition: 
the outer part of the swarm begins to exhibit the milling phenomena 
while the inner part still remains a rigid body. 
Indeed, the collective interaction potential is stronger in the inner 
part of the swarm, and 
particles need a higher kinetic energy injection from the self-driving terms 
to escape the binding potential.
Since the lattice formation of the rigid-body rotation has an ordered
particle distribution, and the milling formation exhibits more disordered
distribution, we propose an {\em ordering factor\/} of period $Q$
to quantitatively distinguish these two states
\begin{equation}
O_{(Q)} \equiv \frac{1}{N \mu} \left| \sum_{i=1}^{N}
{\sum_{j}^{\mu}
{\cos \left( Q \cdot \phi_{j,j+1}^{(i)} \right) } } \right|,
\label{eq:ordering-factor}
\end{equation}
where $\phi_{j ,j+1}^{(i)}$ is the angle between $\vec{x}_{i,j}$
and $\vec{x}_{i,j+1}$ with 
$\vec{x}_{i,j}$ defined as $\vec{x}_j - \vec{x}_i$.
The summation index $j$ here represents the $j$-th nearest neighbor of 
particle $i$, and $\mu$ denotes the number of neighbors that are taken 
into consideration for each particle.
We also define $\vec{x}_{i, \mu +1} \equiv \vec{x}_{i,1}$ to simplify
the formula.
If all $\phi_{j ,j+1}^{(i)}$ are distributed at $2\pi k/Q$ where 
$k<Q$ is a positive integer, $O_{(Q)}=1$, 
and the particles are distributed on a lattice of period $Q$. 
On the other hand, if the distribution is completely random, 
cancellation occurs in the summation of cosines, 
and $O_{(Q)} \simeq 0$ for all $Q$. 
The number of nearest neighbors of each particle $i$ can be arbitrarily 
chosen for $\mu \ge 2$. 
However, note that $\mu$ cannot be too large; otherwise, 
second layer neighbors may be counted, which results in an incorrect $Q$.
For the sake of definiteness, we choose $\mu =3$. 
In order to avoid incorrect estimations due to the dispersed state, 
we also impose that a particle pair must be separated by a distance
no larger than $2 \ell_a$ for the particles to qualify as neighbors. 
Figure\,\ref{cap:ordering-factor}\,(b) shows the distribution of 
$\phi_{j ,j+1}^{(i)}$ collected for all $i$ and $j$ on a rigid-body formation. 
Peaks are observed at $k\pi/3$ ($1 \le k \le 5$), indicating that 
the formation is a hexagonal lattice. 
Figure\,\ref{cap:ordering-factor}\,(c) shows $O_{(Q)}$ versus $Q$  
for the same rigid-body formation.
As expected for a hexagonal lattice, the curve peaks at $Q=6$. 
Therefore, $O_{(6)}$ can be used to explore the transition from a 
hexagonal lattice to a non-lattice mill state. 

Using the quantities defined in 
Eqs.\,(\ref{eq:polarity})\,-\,(\ref{eq:ordering-factor}), different
swarming states can be classified. 
Dramatic changes in $P$, $M$, $M_\mathrm{abs}$, and $O_{(Q)}$  
are observed upon modifying specific parameters in the model and 
indicate a change in the swarming state. 
Figure\,\ref{cap:ordering-factor}\,(a) shows the transition from 
lattice to single-mill states for catastrophic swarms 
as $O_{(6)}$ gradually decreases with respect to increasing $\alpha$.
Also shown in the figure are the same quantities for an H-stable swarm; 
note that as $\alpha$ increases, $O_{(6)}$ suddenly drops to zero,
corresponding to the sudden dissolution of the hexagonal 
lattice structure into a dispersed state.
The larger value of $O_{(6)}$ in the H-stable swarm indicates a 
more regular hexagonal lattice formation. 

\begin{figure}
\includegraphics[%
  width=1.0\columnwidth]{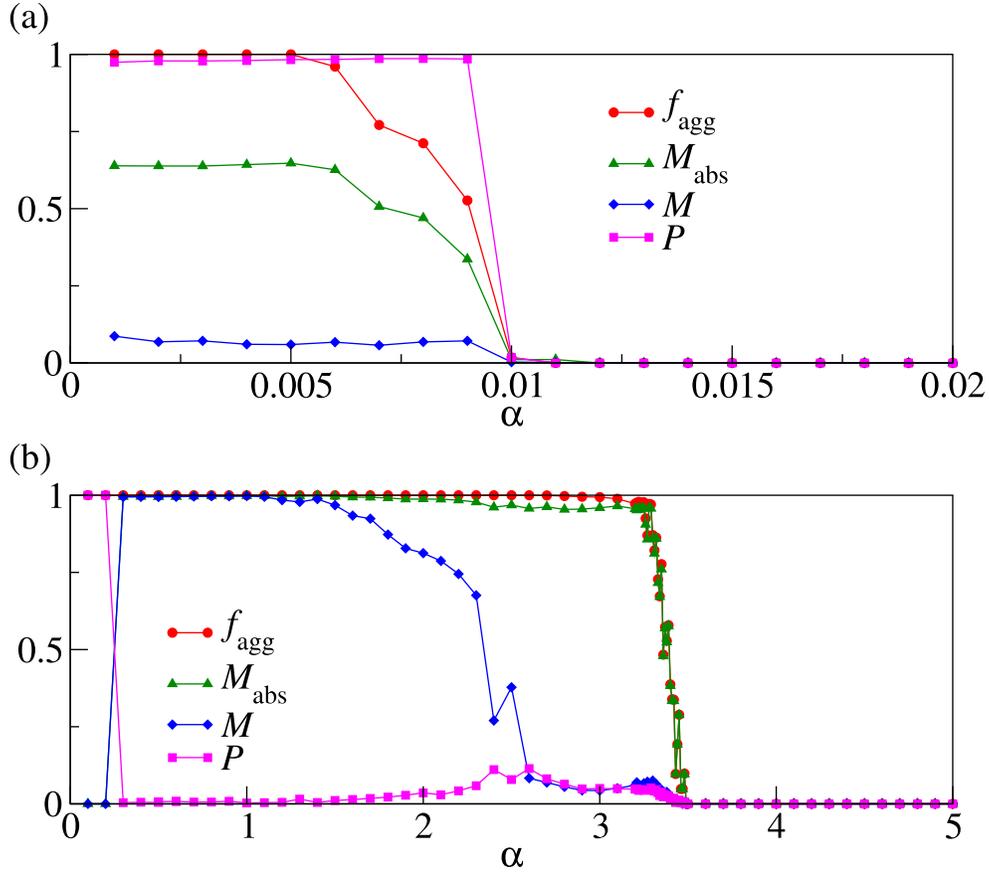}

\caption{\label{cap:transition-measure} The state transition diagram of 
(a) an H-stable swarm and (b) a catastrophic swarm. The fixed parameters
are the same as those in Fig.\,\ref{cap:escap-alpha} with $N=200$.}
\end{figure}

For higher values of $\alpha$, we further consider $P$, $M$, and 
$M_\mathrm{abs}$ to differentiate the coherent state and the two mill states. 
Additionally, in order to distinguish the dispersed state from the rest, 
we calculate the aggregation fraction, $f_\mathrm{agg}$,
defined as the fraction of the $N$ initial particles that aggregate as a swarm.
In Fig.\,\ref{cap:transition-measure}, we show how H-stable swarms
differ from catastrophic ones during the transition between states. 
Figure\,\ref{cap:transition-measure}\,(a) shows that 
the H-stable swarm is a coherent flock for small $\alpha$, 
indicated by $P\simeq1$. 
For increasing $\alpha$, the swarm disperses and $f_\mathrm{agg}=0$.
Note that $P$ remains close to one when $f_\mathrm{agg} \ne 0$,
indicating that the aggregate goes from the coherent lattice state 
directly to the dispersed one. 
Figure\,\ref{cap:transition-measure}\,(b) shows the transition of a 
catastrophic swarm, which displays a full four-stage transition:
in the small $\alpha$ regime, particles arrange as a coherent lattice  
with $P\simeq 1$; as $\alpha$ keeps increasing, the single-mill state appears 
($P\simeq0$ and $M\simeq1$), followed by the double-mill state 
($M_\mathrm{abs}\simeq1$ and $M\simeq0$) until the dispersed state
($f_\mathrm{agg} = 0$) is reached.

Drawing an analogy from the state transition of swarming patterns
to the phase transition of materials, the lattice states can be regarded 
as ``solid''  
since interparticle distances are kept constant. 
The milling state allows particles to ``swim'' within a finite volume 
without being bound to a fixed lattice site; 
thus, it can be regarded as ``liquid''.
Finally, in the dispersed state, particles escape, similarly to 
a ``gas''. 
Upon increasing $\alpha$, 
a catastrophic swarm undergoes the solid--liquid and liquid--gas transitions, 
which resemble the processes of melting and vaporization. 
On the other hand, an H-stable swarm goes from the solid state directly 
to the gas one, which is more similar to sublimation. 

Consistently with granular media models \cite{granular}, 
we may define a ``temperature'' analog
using the variation of the individual particle velocity among the flock: 
$T_\mathrm{s} \equiv \left\langle \left( \vec{v}- \langle \vec{v} 
\rangle \right)^{2} \right\rangle$. 
Note that $\langle \vec{v} \rangle$ is the velocity of the center of mass, 
and thus, $T_\mathrm{s}\simeq0$ for the coherent flock pattern, 
while $T_\mathrm{s} \simeq \alpha/\beta$ for the steady mill states.
The swarming patterns change from one state to another by varying 
$T_\mathrm{s}$. 

While different aggregation morphologies can be studied using
the individual-based model, the large number of degrees of freedom involved 
pose a difficulty for analyzing the dynamics of large $N$ systems. 
In the following section, we therefore develop and investigate a continuum
model consistent with the microscopic description of 
Eqs.\,(\ref{eq:OurPosEq})\,-\,(\ref{eq:interaction_term}).

\section{Continuum model \label{sec:Continuum-model}}
The continuum approach is widely adopted for modeling swarming systems,
especially on the ecological scale where massive movements of populations
are considered \cite{TonerTuSwarming,Mogilner96,BandSolutionSwarming,Mogilner1,TopazSwarming,Topaz2}. 
It is also more suitable for theoretical analysis especially 
in large $N$ limit.
Due to the lack of connections between individual rules and continuum 
``fluxes'', 
most continuum models in the literature are constructed on the basis of 
heuristic arguments. 
Attempts have been made to bridge the gap. 
For example in Ref.\,\cite{Grunbaum1},  
a continuum kinematic one-dimensional advection-diffusion model is 
derived based on a biased random walk process of a set of Poisson-distributed
particles. 
This work is extended to higher dimensions and to more 
general kinematic rules in Ref.\,\cite{Mourelo1,Morale1}.
However, our model is based on dynamic rules and the corresponding 
continuum limit is much more difficult to rigorously justify.
In Ref.\,\cite{FlierlDiscreteModel}, a continuum model is derived 
from a class of dynamic individual-based descriptions by using a 
Fokker-Planck approach.
In order for the flux term to be amenable to analytic investigations, 
the Fokker-Planck equations have to be closed under several assumptions, 
but these assumptions, such as that the preferred velocity which particles 
tend to reach is small with respect to noise terms, 
are not applicable to our model.

In this paper, we derive a continuum model by explicitly calculating the 
ensemble average the model of Eqs.\,(\ref{eq:OurPosEq})\,-\,(\ref{eq:OurMomEq})
using a probability distribution function. 
This classical procedure is described in Ref. \cite{IrvingKirkwood} 
where continuum hydrodynamics equations are derived starting from 
a microscopic collection of $N$ particles. 
Let 
\begin{equation}
f=f\left( \vec{x}_1,\vec{x}_2,...,\vec{x}_N;
\vec{p}_1,\vec{p}_2,...,\vec{p}_N;t \right)
\label{eq:canonical-pdf}
\end{equation}
be the probability distribution function on the phase space,
defined by position and momentum $\left( \vec{x}_i, \vec{p}_i \right)$, 
$1 \le i \le N$, at time $t$. 
The mass density $\rho\left(\vec{x},t\right)$, 
the ensemble velocity field $\vec{u}\left(\vec{x},t\right)$,
and the continuum interaction force $\vec{F}_{V} \left(\vec{x},t\right)$
can be defined as
\begin{eqnarray}
\rho \left( \vec{x},t \right) & = & m \sum_{i=1}^{N}
{\left\langle \delta\left( \vec{x}_i-\vec{x} \right);f \right\rangle },
\label{eq:M-density}\\
\vec{u} \left( \vec{x},t \right) & = & 
\frac{\vec{p} \left( \vec{x},t \right)}{\rho \left( \vec{x},t \right)}
=\frac{\sum_{i=1}^{N}
{\left\langle \vec{p}_i \delta\left(\vec{x}_i-\vec{x}\right);f \right\rangle}}
{\rho \left( \vec{x},t \right)},
\label{eq:continuum velocity}\\
\vec{F}_V \left( \vec{x},t \right) & = & \sum_{i=1}^{N}
{\left\langle -\vec{\nabla}_{\vec{x}_i} U \left( \vec{x}_i \right)
\delta \left( \vec{x}_i-\vec{x} \right);f \right\rangle}.
\label{eq:continuum-force}
\end{eqnarray}
We consider the case of identical masses, $m_i \equiv m$.
The function $\delta \left( \vec{x} \right)$ is the Dirac delta function, and
$U\left( \vec{x}_i \right)$ the collective interaction potential
acting on particle $i$. 
Using the generalized Liouville theorem that incorporates the deformation
of phase space due to the non-Hamiltonian nature of the system at hand
\cite{Tuckerman}, we obtain the continuum equations of motion
\begin{eqnarray}
\frac{\partial \rho}{\partial t} 
+ \vec{\nabla} \cdot \left( \rho\vec{u} \right) & = & 0,
\label{eq:cont-eqn0} \\
\hspace{-5.0mm}
\frac{\partial}{\partial t} \left( \rho \vec{u} \right)
+ \vec{\nabla} \cdot \left( \rho\vec{u}\vec{u} \right) 
+ \vec{\nabla} \cdot \hat{\sigma}_\mathrm{K} & = & 
\alpha \rho \vec{u} - 2 \beta E_\mathrm{K} \vec{u}
-2 \beta \vec{q}_\mathrm{K} + 2 \beta \vec{u} \cdot \hat{\sigma}_\mathrm{K} 
+ \vec{F}_V.
\label{eq:mom-tran-eqn0}
\end{eqnarray}
The first is the equation of continuity, and 
the second is the momentum transport equation.
Here, $E_\mathrm{K} = \rho \left| \vec{u} \right|^{2} /2$ 
is the kinetic energy. 
The terms ${\vec{q}}_\mathrm{K} \left( \vec{x},t \right)$ and 
$\hat{\sigma}_\mathrm{K} \left( \vec{x},t \right)$
are mathematically defined as 
\begin{eqnarray*}
\vec{q}_\mathrm{K} \left( \vec{x},t \right) & = & \sum_{i=1}^{N}
{\left\langle \frac{m}{2} \left| \frac{\vec{p}_i}{m}-\vec{u} \right|^{2}
\left( \frac{\vec{p}_i}{m}-\vec{u} \right) 
\delta \left( \vec{x}_i-\vec{x} \right);f \right\rangle }, 
\\
\hat{\sigma}_\mathrm{K} \left( \vec{x},t \right) & = & \sum_{i=1}^{N}
{m \left\langle \left( \frac{\vec{p}_i}{m}-\vec{u} \right) 
\left( \frac{\vec{p}_i}{m} - \vec{u} \right) 
\delta \left( \vec{x}_i-\vec{x} \right);f \right\rangle },
\end{eqnarray*}
and represent the energy flux and the stress tensor due to local 
fluctuations in particle velocities with respect to 
$\vec{u} \left( \vec{x},t \right)$.
The derivation of the term $\vec{\nabla} \cdot \hat{\sigma}_\mathrm{K}$ 
can be found in Ref. \cite{IrvingKirkwood} while the other terms related to 
$\vec{q}_\mathrm{K}$ and $\hat{\sigma}_\mathrm{K}$ are derived in Appendix A. 
By simulating the discrete model, we estimate the magnitude of 
$\vec{q}_\mathrm{K}$ and $\hat{\sigma}_\mathrm{K}$ and find that both 
fluctuation terms become negligible with respect to the the other terms 
on the RHS of Eq.\,(\ref{eq:mom-tran-eqn0}) in the lattice, single-mill, 
and the dispersed states. 
Thus, neglecting the fluctuation terms, we obtain 
\begin{eqnarray}
\frac{\partial\rho}{\partial t} 
+ \vec{\nabla} \cdot \left( \rho\vec{u} \right) & = & 0,
\label{eq:continuum-model-1}\\
\frac{\partial}{\partial t} \left( \rho\vec{u} \right) 
+ \vec{\nabla} \cdot \left( \rho\vec{u}\vec{u} \right) & = & 
\alpha \rho \vec{u} - 2 \beta E_\mathrm{K} \vec{u} + \vec{F}_V.
\label{eq:continuum-model-2}
\end{eqnarray}

\subsection*{Continuum interaction force}

In Eq. (\ref{eq:continuum-model-2}), the continuum interaction force can be 
obtained by substituting the explicit form of the interaction potential 
Eq.\,(\ref{eq:interaction_term}) into Eq.\,(\ref{eq:continuum-force})
\begin{equation}
\vec{F}_V \left( \vec{x},t \right) = \sum_{i=1}^{N} \sum_{j=1}^{N}
{\left\langle -\vec{\nabla}_{\vec{x}_{i}} V \left( \vec{x}_i-\vec{x}_j \right)
\delta \left( \vec{x}_i - \vec{x} \right);f \right\rangle }.
\label{eq:step-1}
\end{equation}
Using the fact that an arbitrary function $F\left(\vec{x}_j\right)$ 
$\forall \vec{x}_j \in \Rset^d$ can be written as
\[
F \left( \vec{x}_j \right) = \int_{\Rset^d}{F \left( \vec{y} \right) 
\delta \left( \vec{x}_j-\vec{y} \right)} \mathrm{d}\vec{y},
\]
we can rewrite Eq.\,(\ref{eq:step-1}) as
\begin{eqnarray}
\vec{F}_V \left( \vec{x},t \right) & = & \sum_{i=1}^{N}\sum_{j=1}^{N}
{\int_{\Rset^d} \mathrm{d}\vec{y}{ \left\langle -\vec{\nabla}_{\vec{x}_i}
V \left( \vec{x}_i-\vec{y} \right)  
\delta \left( \vec{x}_i-\vec{x} \right) 
\delta \left( \vec{x}_j-\vec{y} \right);f \right\rangle }} 
\nonumber \\
& = & \int_{\Rset^d}
{-\vec{\nabla}_{\vec{x}} V \left( \vec{x}-\vec{y} \right)}
\sum_{i=1}^{N}\sum_{j=1}^{N}
{\left\langle \delta \left( \vec{x}_j-\vec{y} \right) 
\delta \left( \vec{x}_i-\vec{x} \right);f \right\rangle }\mathrm{d}\vec{y}
\nonumber \\
 & = & \int_{\Rset^d}{-\vec{\nabla}_{\vec{x}} V \left( \vec{x}-\vec{y} \right)
\rho^{(2)} \left( \vec{x},\vec{y},t \right) \mathrm{d}\vec{y},}
\label{eq:step-4}
\end{eqnarray}
where the $\rho^{(2)}$ is the pair density
\[
\rho^{(2)} \left( \vec{x},\vec{y},t \right) \equiv \sum_{i=1}^{N}\sum_{j=1}^{N}
{\left\langle \delta \left( \vec{x}_j-\vec{y} \right) 
\delta \left( \vec{x}_i-\vec{x} \right);f \right\rangle }.
\]
Note that we should take the ensemble average on a scale 
considerably larger than the spacing between particles.
If the particles are quite dispersed, the suitable scale may be much larger 
than the characteristic lengths of the interaction force ($-\vec{\nabla} V$ in 
Eq.\,(\ref{eq:step-4})), rendering it localized.
In this case, the continuum approach cannot capture the swarming 
characteristics occurring on the the interaction scale and fails 
to describe the individual-based model on such a scale.
This is what occurs in the H-stable regime, which we further discuss in 
Sec.\,\ref{sec:comparison}.

For identical particles, 
the pair density can be written as
\[
\rho^{(2)} \left( \vec{x},\vec{y},t \right) = \frac{1}{m^2}
\rho \left( \vec{x},t \right) \rho \left( \vec{y},t \right)
g^{(2)} \left( \vec{x},\vec{y} \right),
\]
where the correlation function $g^{(2)} \left( \vec{x},\vec{y} \right) = 1$ 
when the particles 
have no intrinsic correlation. 
Using this assumption,
\begin{equation}
\rho^{(2)} \left( \vec{x},\vec{y},t \right) = \frac{1}{m^2}
\rho \left( \vec{x},t \right) \rho \left( \vec{y},t \right),
\label{eq:pair-density}
\end{equation}
and
\begin{eqnarray}
\vec{F}_V \left( \vec{x},t \right) & = & \int_{\Rset^d}
{-\vec{\nabla}_{\vec{x}} V \left( \vec{x}-\vec{y} \right) 
\frac{1}{m^2} \rho \left( \vec{x},t \right) 
\rho \left( \vec{y},t \right)} \mathrm{d} \vec{y}.
\label{eq:explicit-continuum-force}
\end{eqnarray}
If we further substitute the interaction potential specified in 
Eq.\,(\ref{eq:interaction_term}) into the above equation, we get
\begin{eqnarray}
\vec{F}_V \left( \vec{x},t \right) & = & -\rho \left( \vec{x},t \right)
\vec{\nabla} \int_{\Rset^d}{\left( 
-\frac{C_a}{m^2} \e^{-\frac{\left| \vec{x}-\vec{y} \right|}{\ell_a}}
+\frac{C_r}{m^2} \e^{-\frac{\left| \vec{x}-\vec{y} \right|}{\ell_r}}
\right) \rho \left( \vec{y},t \right) } \mathrm{d}\vec{y}.
\label{eq:Lavine-continuum-force}
\end{eqnarray}

Since we assume that all particles have an identical mass, we may choose
$m=1$ without loss of generality. 
In this case, Eq.\,(\ref{eq:Lavine-continuum-force}) becomes 
the one proposed in Ref.\,\cite{LevineModel}, on heuristic grounds. 
Using Eq.\,(\ref{eq:continuum-model-1}), 
we may modify Eq.\,(\ref{eq:continuum-model-2}) and divide by $\rho$ on both
sides to obtain a more conventional expression 
\begin{eqnarray}
\frac{\partial\rho}{\partial t} 
+ \vec{ \nabla} \cdot \left( \rho\vec{u} \right) & = & 0, 
\label{eq:continuum-model-3} \\
\frac{\partial \vec{u}}{\partial t} + \vec{u} \cdot \nabla \vec{u} & = & 
\alpha \vec{u} - \beta \left| \vec{u} \right|^{2} \vec{u}
- \frac{1}{m^2} \vec{\nabla} \int_{\Rset^d}
{ V \left( \vec{x}-\vec{y} \right) 
\rho \left( \vec{y},t \right)} \mathrm{d} \vec{y}.
\label{eq:continuum-model-4}
\end{eqnarray}

\section{Comparison to the individual-based model \label{sec:comparison}}
\begin{figure}
\includegraphics[%
  width=1.0\columnwidth]{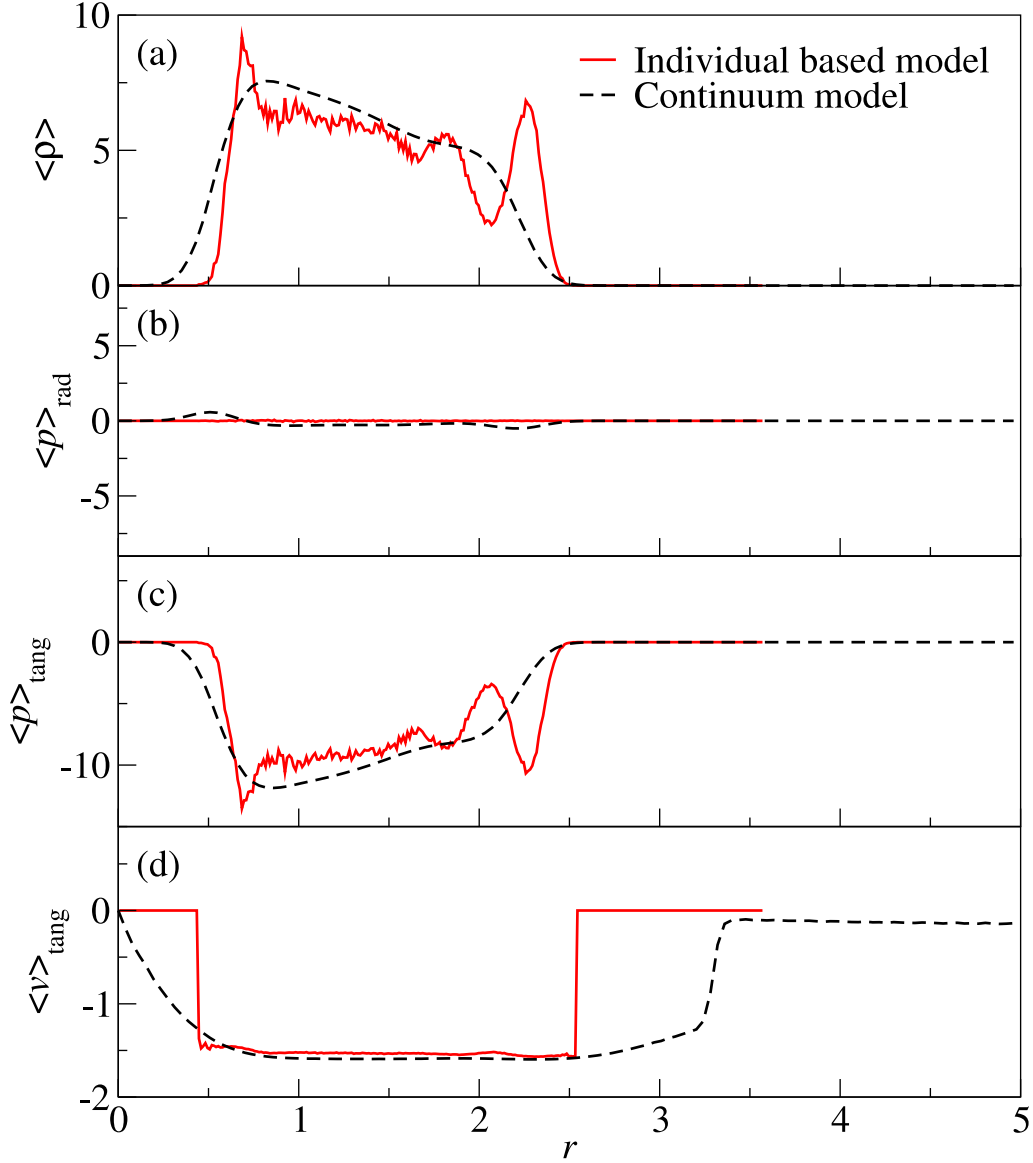}

\caption{\label{cap:continuum-discrete-compare}Comparison of the numerical
simulations of the individual-based model and the continuum model:
The parameters used in both simulations are $C_a=0.5$, $C_r=1.0$, 
$\ell_a=2.0$, $\ell_r=0.5$, $\alpha=1.2$, $\beta=0.5$, and the total
mass $m_\mathrm{tot}=88$. (a) The averaged density profiles along the 
radial distance from the center of mass. (b) The averaged radial momentum
profiles. (c) The averaged tangential momentum profiles. (d) The averaged 
tangential velocity profiles. }
\end{figure}

The time-dependent variations of the density $\rho \left( \vec{x},t \right)$ 
and of the momentum $\vec{p} \left( \vec{x},t \right) \equiv \rho \left( \vec{x},t \right) \vec{u}\left( \vec{x},t \right)$ 
can be obtained through numerical simulations of 
Eqs.\,(\ref{eq:continuum-model-1})\,-\,(\ref{eq:continuum-model-2}).
Here, we use the Lax-Friedrichs method \cite{Leveque:num-pde} 
to integrate the partial differential equations.
We compare the results of the continuum model to those of 
the individual-based model of 
Eqs.\,(\ref{eq:OurPosEq})\,-\,(\ref{eq:OurMomEq}).
Figure\,\ref{cap:continuum-discrete-compare} shows the frequently observed 
single-mill steady state solutions of both models in the catastrophic regime. 
Both simulations use identical parameter values and the same total mass 
\begin{eqnarray*}
m_\mathrm{tot} & = & 
\int_{\infty}{\rho\left( \vec{x} \right)} \mathrm{d} \vec{x} 
 = Nm,
\end{eqnarray*}
which are listed in the figure caption.
In the individual-based model, particles are initially distributed with
random velocities and at random positions in a $2\ell_a \times 2\ell_a$ box.
The initial condition of the continuum model is a homogeneous 
density in a box of the same size with randomized momentum field. 
While the individual-based model adopts a free boundary condition 
that allows the particles to move around over the entire space, 
the continuum model uses an equivalent boundary condition of out-going waves
but on a fixed computational domain of $5\ell_a \times 5\ell_a$ size which is
divided into $256 \times 256$ grid cells.
The individual-based simulation uses an adaptive time step size that 
keeps the increment in position of each step under $\ell_a / 10$ and 
the increment in velocity under $C_a / 5 m$.
The time step size of the continuum simulation is chosen so that the 
CFL number is $0.98$.
Figure\,\ref{cap:continuum-discrete-compare}\,(a) illustrates the averaged 
density $\langle \rho \rangle$ as a function of the radial distance 
from the center of mass.
These two profiles are in good agreement despite the density 
oscillation
shown in the individual-based model, reflecting a multiple-ring 
ordering of the particle distribution.
Figures\,\ref{cap:continuum-discrete-compare}\,(b) and (c) match the averaged 
radial and tangential momenta (denoted by $\langle p \rangle_\mathrm{rad}$ 
and $\langle p \rangle_\mathrm{tang}$ respectively) from the simulations of 
both models.
The negligible radial momenta in 
Fig.\,\ref{cap:continuum-discrete-compare}\,(b) indicate that there is no 
net inward or outward mass movement, 
and thus, the density profile along the radial direction is steady.
We can divide the momentum by the density to obtain the velocity field.
In Fig.\,\ref{cap:continuum-discrete-compare}\,(d), we show the averaged 
tangential velocities, $\langle v \rangle _\mathrm{tang} \equiv 
\langle p \rangle_\mathrm{tang} / \langle \rho \rangle$, from the simulations 
of both models;
it shows that both the individual-based and the continuum swarms are 
rotating at the same constant speed, which equals to $v_\mathrm{eq}$.
%

\subsection*{Validity of the continuum model}

\begin{figure}
\includegraphics[%
  width=1.0\columnwidth]{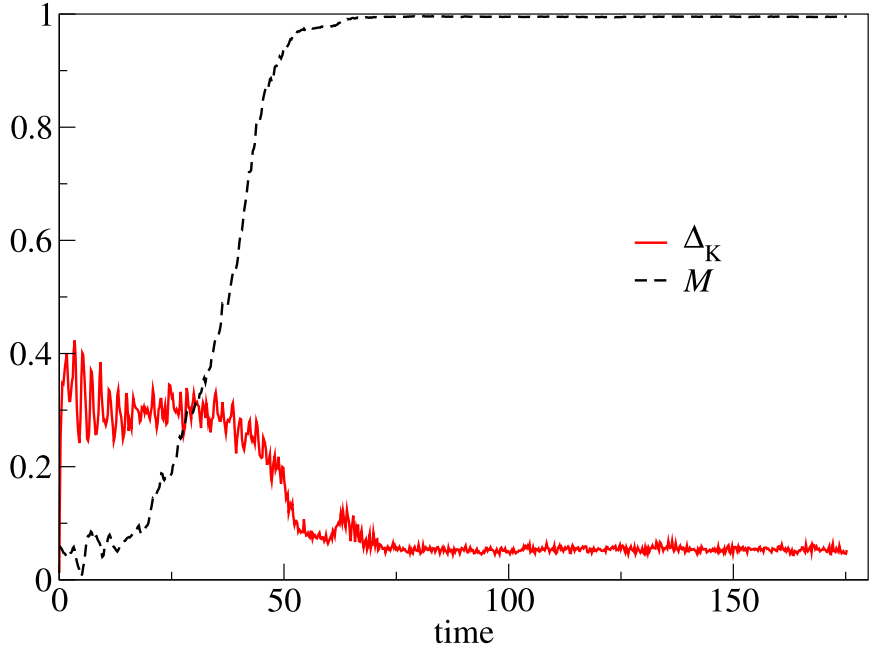}

\caption{\label{cap:speed-variation} Relative speed fluctuations 
$\Delta_\mathrm{K}$ while forming a single mill from random initial conditions.
The dashed curve illustrates the normalized angular momentum $M$ in 
Eq.\,(\ref{eq:ang-momen}) while the solid curve represents $\Delta_\mathrm{K}$.
The parameters of this simulation are $\alpha=1.0$, $\beta=0.5$, 
$C_a=0.5$, $C_r=1.0$, $\ell_a=2.0$, $\ell_r=0.5$, and $N=500$. }
\end{figure}

The ensemble average implicit in the continuum approach does not allow for 
double-milling in the continuum limit  
because the velocity inside a mesh cell is averaged and unified. 
Local velocity variations, which contribute to 
$\vec{q}_\mathrm{K} \left( \vec{x},t \right)$ and 
$\hat{\sigma}_\mathrm{K} \left( \vec{x},t \right)$ 
in Eq.\,(\ref{eq:mom-tran-eqn0}), 
are neglected.  
We calculate the ratio of the speed variation to the equilibrium speed,  
$\Delta_\mathrm{K} \equiv \sqrt{ \langle \left( v - v_\mathrm{eq} \right)^2 
\rangle / {v_\mathrm{eq}}^2}$, in order to efficiently estimate the 
contribution of these local velocity fluctuation terms.
Figure\,\ref{cap:speed-variation} shows that $\Delta_\mathrm{K}$
becomes negligible after the swarm has reached the 
single-mill configuration and $M\simeq1$. 
However, during the transient time, $\Delta_\mathrm{K}$ is significantly 
larger, which implies that $\vec{q}_\mathrm{K} \left( \vec{x},t \right)$ and 
$\hat{\sigma}_\mathrm{K} \left( \vec{x},t \right)$ cannot be neglected
during this period. 
Hence, the continuum model of 
Eqs.\,(\ref{eq:continuum-model-1})\,-\,(\ref{eq:continuum-model-2}) 
can be useful in analyzing the stability of the steady state solution 
but does not capture the dynamics of the swarm settling into this steady state.

\begin{figure}
\includegraphics[%
  width=1.0\columnwidth]{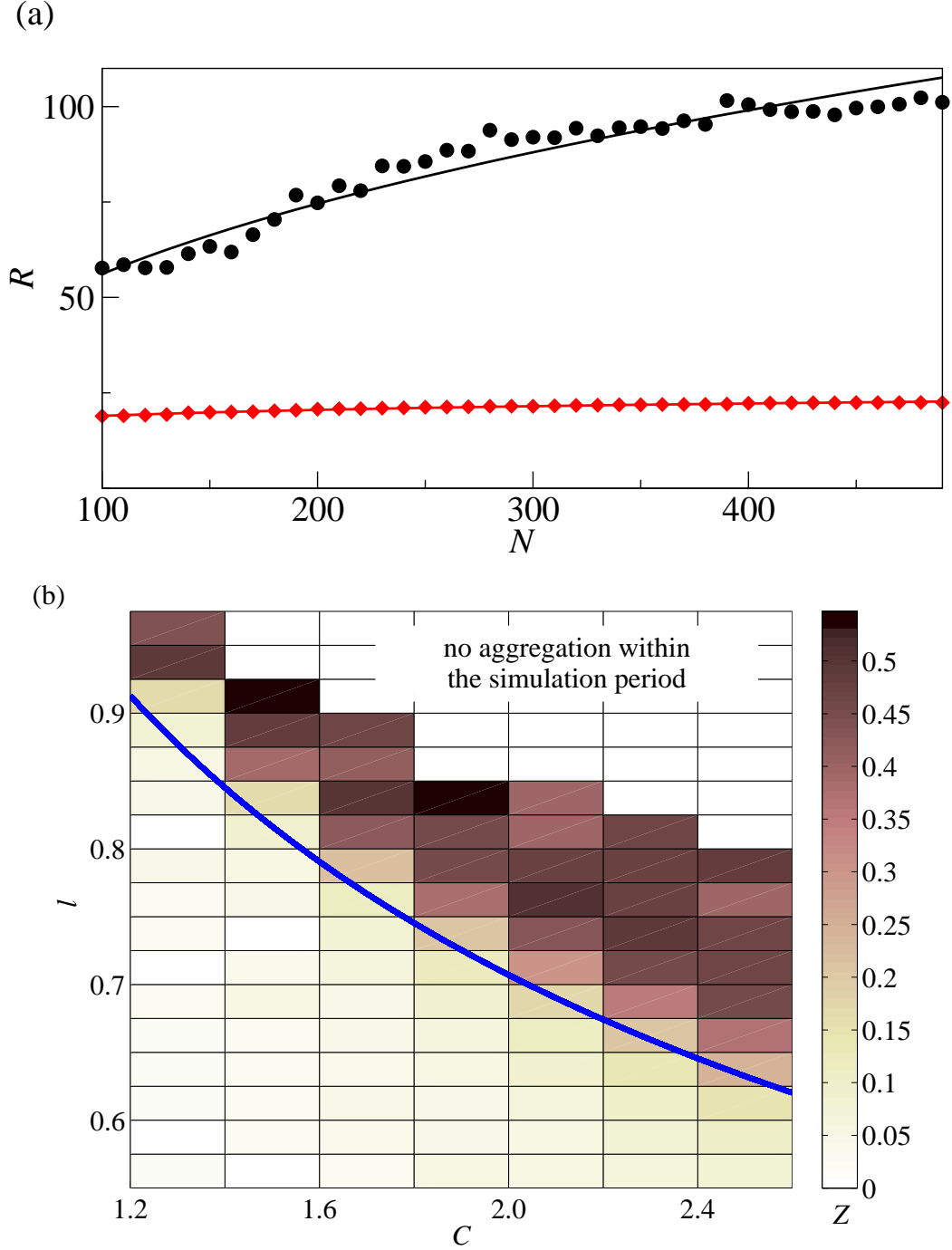}

\caption{\label{cap:rigid-radius}(a) Flock radius versus number of particles
with the total mass fixed at $m_\mathrm{tot}=Nm=500$. Solid circles 
represent the H-stable flock ($\ell_r=1.5$), fitted by $R \propto N^{0.41}$,  
while the solid diamonds represent the catastrophic flock ($\ell_r=1.3$),
fitted by $R \propto N^{0.11}$. 
The other parameters are $\alpha=0$, $\beta=0.5$, $C_a=0.5$, 
$C_r=1.0$, and $\ell_a=2.0$.
(b) Exponents $Z$ of the power law fitting $R \propto N^Z$ for a range
of $C$ and $\ell$. The dimensionless parameters $C$ and $\ell$ are changed
by varying $C_r$ and $\ell_r$ while the other parameters remain the same as 
above. The color map indicates the power $Z$, and darker colors represent
higher exponents. The solid curve marks the boundary between the H-stable 
and the catastrophic regions. }
\end{figure}

While Figure\,\ref{cap:continuum-discrete-compare} shows good agreement
between the steady state solutions of the continuum and the individual-based
models in the catastrophic regime, inconsistencies arise as the parameters 
shift into the H-stable regime. 
Here, at low particle speeds, 
the individual-based model results in compactly supported solutions 
similar to those shown in Fig.\,\ref{cap:hstable-mill}.
Conversely, the corresponding continuum model yields 
a uniform density distribution spreading over the entire computational 
domain regardless of domain size.
Since a valid continuum model should reflect the large number limit of
the individual-based model, we can investigate how the steady state solutions
of the individual-based model evolve by increasing $N$ while keeping all 
continuum variables and parameters fixed.
In particular, we increase $N$ while keeping the macroscopic parameter 
$m_\mathrm{tot} = N m$ fixed.
If the individual-based solution has converged to the continuum limit,
the solutions should be independent of any microscopic parameter,
such as $N$.
In Fig.\,\ref{cap:rigid-radius}\,(a), we show the radius $R$ of steady swarms  
versus $N$ under fixed $m_\mathrm{tot}$ for an H-stable case  
and for a catastrophic one.
The flock size is indeed independent of $N$ for catastrophic swarms, 
and the two models yield consistent solutions, as shown in 
Fig.\,\ref{cap:continuum-discrete-compare}.
However, in the H-stable regime, the swarm size increases with
$\sqrt{N}$ in spite of a fixed $m_\mathrm{tot}$.
This suggests that a compactly supported solution does not exist
in the large number limit of an H-stable swarm.
Figure\,\ref{cap:rigid-radius}\,(b) further illustrates this point by
expanding the investigation to a broader parameter space. 
The H-stability threshold is the solid curve, parting the $C$--$\ell$ 
phase space in Fig.\,\ref{cap:rigid-radius}\,(b).
The flock radius $R$ is approximately independent of $N$ for 
catastrophic swarms, 
but when the parameters $C$ and $\ell$ cross over to the H-stable regime, 
$R$ scales as $N^Z$ with $Z \simeq 1/2$. 
As $N \rightarrow \infty$, H-stable swarms tend to occupy the entire space.

\begin{figure}
\includegraphics[%
  width=1.0\columnwidth]{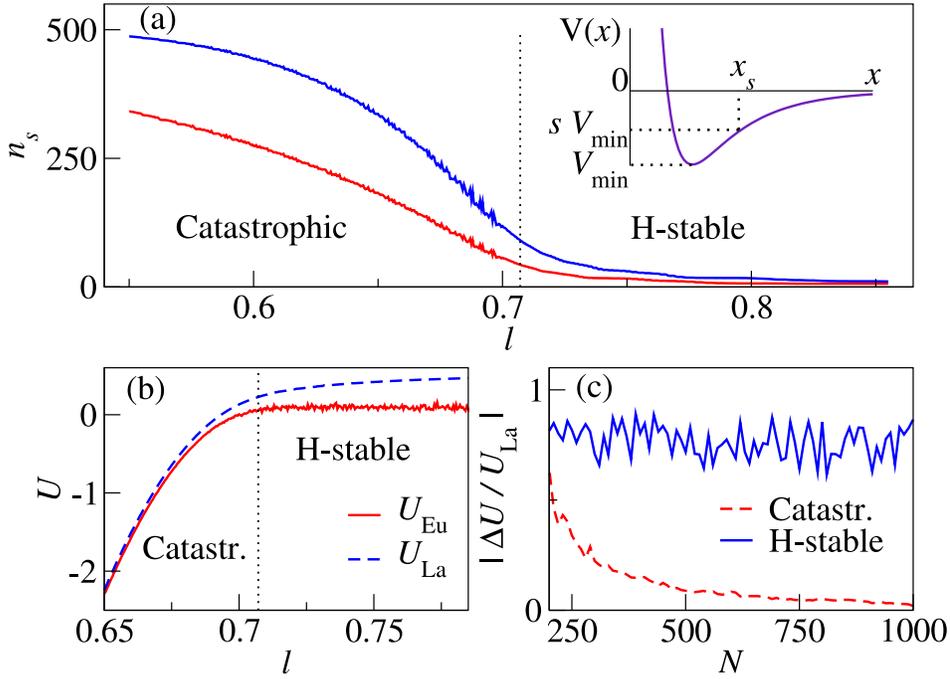}

\caption{\label{cap:sig-neighbors}(a) $n_s$ versus $\ell$. The upper curve
represents the case of $s = 10\%$ while the lower curve is for $s = 50\%$.
On the upper-right corner is an illustration showing how the significant
neighbors are defined.
(b) $U_\mathrm{Eu}$ and $U_\mathrm{La}$ versus $\ell$. 
Here $\alpha=0.003$, $\beta=0.5$, $C_a = 0.5$, $\ell_a = 2.0$, $C_r = 1.0$,
and $N=500$.
(c) $\triangle U$ and versus $N$. $\ell = 0.65$ for the catastrophic curve 
while $\ell=0.75$ for the H-stable curve. The other parameter values are
the same as (b). }
\end{figure}

The cue to the inconsistency between the solutions of the two models in 
the H-stable regime lies in the derivation of the continuum model. 
As previously mentioned, 
the macroscopic variables are obtained as ensemble averages over a large
number of microscopic ones. 
In the catastrophic regime, $\delta_\mathrm{NND} \ll \ell_a,\,\ell_r$ 
in large $N$ limit, as shown in Fig.\,\ref{cap:rigid-radius}.
Hence, as $N \rightarrow \infty$, the particle distribution converges to a 
continuum density on a scale comparable to the interaction length.
On the other hand, for an H-stable swarm, $\delta_\mathrm{NND}$ stays 
non-negligible with respect to the characteristic length of the interaction.
Hence, Eq.\,(\ref{eq:step-4}) does not hold on a scale comparable to the
interaction length, and as a result, 
Eq.\,(\ref{eq:explicit-continuum-force}) is not a valid description of the 
continuum force on such a scale in the H-stable regime. 
This is also verified in Fig.\,\ref{cap:sig-neighbors}.
In Fig.\,\ref{cap:sig-neighbors}\,(a), we define significant neighbors of 
a particle as neighbors that exhibit a ``significant interaction''. 
The quantitative definition is illustrated by the graph on the upper-right 
corner, in which the pairwise interaction potential $V \left( r \right)$ is 
plotted versus the inter-particle distance $r$, and 
the potential well depth is denoted by $V_\mathrm{min}$.
We define a distance $r_s$ so that $V \left( x \right) > s V_\mathrm{min}$ 
if $x > x_s$, where $0 \le s \le 1$ is a ratio.
Then, the number of significant neighbors of each particle is the number
of neighbors at a distance $x$ for which $x \le x_s$. 
In Fig.\,\ref{cap:sig-neighbors}\,(a), we count the averaged number of 
significant neighbors of a particle, denoted by $n_s$, for $s=0.5$ and 
$s=0.1$. 
In the H-stable regime, $n_s$ is very low and remains steady; 
it rises rapidly when the parameter crosses over into the catastrophic regime. 
The results suggest that the H-stable swarms are locally too sparse  
for Eq.\,(\ref{eq:step-4}) (and thus, Eq.\,(\ref{eq:explicit-continuum-force}))
to remain valid.
Furthermore, we can use ensemble averages to approximate the collective 
interaction potentials in the two models.
If the continuum limit properly describe the individual-based description, 
these two potential energies should converge as $N$ increases.
Let us define $U_\mathrm{Eu}$ as the continuum ensemble average interaction 
potential in the Eulerian frame
\begin{eqnarray}
U_\mathrm{Eu} \left( \vec{x} \right) & \equiv & 
\frac{1}{m^2} \int_{\Rset^d} 
{ V \left( \vec{x}-\vec{y} \right) \rho \left( \vec{x} \right) 
\rho \left( \vec{y} \right)}\,\mathrm{d} \vec{y}.
\nonumber
\end{eqnarray}
Here $\rho \left( \vec{x} \right)$ is approximated by the ensemble average of 
the individual particles during the simulation.
We also define $U_\mathrm{La}$ as the average collective potential calculated 
in the Lagrangian frame, 
$U_\mathrm{La} \left( \vec{x} \right) \equiv \langle U \left( \vec{x}_i \right) \rangle_{\vec{x}_i=\vec{x}}$, where 
$U \left( \vec{x}_i \right)$ is defined in Eq.\,(\ref{eq:interaction_term}) 
of the individual-based model.
Since the rigid-body rotation and the single-mill state have rotational 
symmetry with respect to $\vec{x}_\mathrm{CM}$, 
we evaluate $U_\mathrm{Eu}$ and $U_\mathrm{La}$ after such states are reached
and at position $\vec{x}$ such that 
$\left| \vec{x} - \vec{x}_\mathrm{CM} \right| = R/2$,
where $R$ is the swarm radius. 
In Fig.\,\ref{cap:sig-neighbors}\,(b), $U_\mathrm{Eu}$ and $U_\mathrm{La}$ 
are shown to converge in the catastrophic regime and diverge in the H-stable 
one.
In Fig.\,\ref{cap:sig-neighbors}\,(c), we investigate whether the difference 
between these two averaged potentials, 
$\triangle U \equiv U_\mathrm{Eu} - U_\mathrm{La}$, vanishes with increasing 
$N$.
Figure\,\ref{cap:sig-neighbors}\,(c) shows that $\triangle U$ indeed tends 
to zero for catastrophic swarms by increasing $N$ but remains finite for 
H-stable ones. 
For the ensemble average to be valid in the H-stable regime, we may instead 
choose a scale that is much larger than the characteristic interaction lengths.
Under such low resolution, the particle distribution can be seen as
a continuous density, and Eq.\,(\ref{eq:step-4}) is then valid.
This becomes the case of the incompressible fluids in 
Ref.\,\cite{IrvingKirkwood}, where the interaction is extremely localized,
and hence, the continuum force yields a stress tensor as a function of the 
local density. 
However, the swarming patterns which we are interested in emerge on a much 
smaller scale.
In contrast, when the particles are in the dispersed state, they are far
away from each other;
thus, particle-particle interaction is very weak and dominated by velocity
fluctuations. 
The continuum force then yields a scalar pressure, 
which gives the gas dynamics equations \cite{IrvingKirkwood}.

\section{Stability of the homogeneous solution \label{sec:linear-stability}}
That the solutions of the continuum model of Eqs. (\ref{eq:continuum-model-1})
and (\ref{eq:continuum-model-2}) relax toward a uniform density 
distribution in the H-stable regime can also be shown by the 
linear stability analysis of its homogeneous solution. 
Let us first consider a more general case for a 2D self-driving continuum model
with a non-local interaction 
\begin{eqnarray}
\frac{\partial \rho}{\partial t} 
- \vec{\nabla} \cdot \left( \rho \vec{u} \right) & = & 0;
\label{eq:continuum-model-5} \\
\frac{\partial}{\partial t} \left( \rho \vec{u} \right) 
+ \vec{\nabla} \cdot \left( \rho \vec{u} \vec{u} \right) & = & 
f \left( \left| \vec{u} \right| \right) \rho \vec{u} 
- \rho \vec{\nabla} \int_{\Rset^d} { 
V \left(\left| \vec{x} - \vec{y} \right|\right)
\rho \left(\vec{y}\right) 
}\,\mathrm{d} \vec{y},
\nonumber 
\end{eqnarray}
where $f\left( \left| \vec{u} \right| \right)$ is a scalar function specifying
the self-driving mechanism, and the non-local interaction is expressed
by the convolution term. 
For our model, 
$
f \left( \left| \vec{u} \right| \right) = 
\alpha - \beta \left| \vec{u} \right|^{2}
$.
The possible homogeneous steady state solutions can be written as
$\rho \left( \vec{x},t\right ) = \rho_0$ and 
$\vec{u} \left( \vec{x},t \right) = v_0\,\hat{v}$,
where $\hat{v}$ is a unit vector and $v_0$ can be $0$ or any of the
roots of $f \left( v_0 \right) = 0$. 
For our Rayleigh-type dissipation, $v_0 = \sqrt{\alpha/\beta}$.
We perturb the steady state solution using 
$
\rho \left( \vec{x},t \right) = \rho_0 
+ \delta \rho \exp \left( \sigma t + \mathrm{i} \vec{q} \cdot \vec{x} \right)$
and $\vec{u} \left( \vec{x},t \right) = v_0\,\hat{v} 
+ \left( \delta u\,\hat{u} + \delta v\,\hat{v} \right) 
\exp \left( \sigma t + \mathrm{i} \vec{q} \cdot \vec{x} \right)
$,
where $\delta \rho$, $\delta u$, $\delta v\ll1$ are small amplitudes.
The unit vector $\hat{u}$ points to the direction perpendicular to 
$\hat{v}$ on the 2D space. 
The wave vector is denoted by $\vec{q}$ while 
$\sigma = \sigma \left( \vec{q} \right)$ represents its growth rate. 
By substituting this ansatz into 
Eq.\,(\ref{eq:continuum-model-5}), 
the dispersion relation is
\begin{eqnarray}
\sigma^{\prime} \left( 
   \begin{array}{c}
   \delta\rho\\
   \delta u\\
   \delta v
   \end{array}
\right) & = & \left(
   \begin{array}{ccc}
   0 & 
   -\mathrm{i} \rho_0 q \sin\theta &
   -\mathrm{i} \rho_0 q \cos\theta \\
   -\mathrm{i} q \tilde{V} \left( \vec{q} \right) \sin\theta & 
   f \left( v_0 \right) &
   0 \\
   -\mathrm{i} q \tilde{V} \left( \vec{q} \right) \cos\theta & 
   0 &
   f \left( v_0 \right) + v_0 f^{\prime} \left( v_0 \right)
   \end{array}
\right) \left(
   \begin{array}{c}
   \delta \rho \\
   \delta u \\
   \delta v
   \end{array}
\right),
\label{eq:dispersion-relation-matrix}
\end{eqnarray}
where $\sigma^{\prime} \equiv \sigma + \mathrm{i} v_0\,\hat{v} \cdot \vec{q}$, 
and $\tilde{V} \left( \vec{q} \right)$ is the Fourier transform of the 
pairwise interaction potential $V \left( \vec{x} \right)$.
The angle between the wave vector $\vec{q}$ and the unit vector $\hat{u}$
is denoted by $\theta$.

For the case of $v_0=0$, the solution is isotropic, and we can
arbitrarily choose the unit vector $\hat{v}$. If the wave vector
$\vec{q}$ is parallel to the arbitrarily chosen $\hat{v}$, 
Eq.\,(\ref{eq:dispersion-relation-matrix}) reduces to
\begin{eqnarray*}
\sigma \left( 
   \begin{array}{c}
   \delta \rho\\
   \delta u\\
   \delta v
   \end{array}
\right) & = & \left( 
   \begin{array}{ccc}
   0 & 
   0 & 
   - \mathrm{i} \rho_0 q \\
   0 & 
   f \left( 0 \right) & 
   0 \\
   \mathrm{i} q \tilde{V} \left( \vec{q} \right) & 
   0 & 
   f \left( 0 \right)
   \end{array}
\right) \left( 
   \begin{array}{c}
   \delta \rho \\
   \delta u \\
   \delta v
   \end{array}
\right),
\end{eqnarray*}
and $\sigma=f \left( 0 \right)$ or ${\left( f \left(0 \right) 
\pm \sqrt{f \left( 0 \right)^{2} 
- 4\rho_0 q^2 \tilde{V} \left( \vec{q} \right)} \right)} / 2$.
If $f \left( 0 \right) > 0$, the homogeneous solution is always unstable.
If $f \left( 0 \right) < 0$, the homogeneous solution is stable only
when $\rho_0 q^2 \tilde{V} \left( \vec{q} \right) > 0$. 
Since $\rho_0$ and $q^2$ are both non-negative, the criterion can be reduced to
\begin{equation}
\tilde{V} \left( \vec{q} \right) > 0.
\label{eq:linear-stable-criterion}
\end{equation}
For our Rayleigh-type dissipation, $f \left( 0 \right) = \alpha$.
Since $\alpha$ is positive, the uniform density solution
with zero speed is an unstable steady state solution.

For the case of $v_0 \ne 0$ satisfying $f \left( v_0 \right) = 0$, 
Eq.\,(\ref{eq:dispersion-relation-matrix}) becomes
\begin{eqnarray*}
\sigma^{\prime} \left( 
   \begin{array}{c}
   \delta \rho \\
   \delta u \\
   \delta v
   \end{array}
\right) = &  & \left( 
   \begin{array}{ccc}
   0 & 
   - \mathrm{i} \rho_0 q \sin\theta &
   - \mathrm{i} \rho_0 q \cos\theta \\
   - \mathrm{i} q \tilde{V} \left( \vec{q} \right) \sin\theta & 
   0 &
   0 \\
   - \mathrm{i} q \tilde{V} \left( \vec{q} \right) \cos\theta & 
   0 &
   v_0 f^{\prime} \left( v_0 \right)
   \end{array}
\right) \left( 
   \begin{array}{c}
   \delta \rho \\
   \delta u \\
   \delta v
   \end{array}
\right).
\end{eqnarray*}
We thus obtain the growth rate by solving the following eigenvalue equation 
\begin{eqnarray}
{\sigma^{\prime}}^{3} 
- v_0 f^{\prime} \left( v_0 \right) {\sigma^{\prime}}^{2}
+ \sigma^{\prime} \rho_0 q^2 \tilde{V} \left( \vec{q} \right)
-v_0 f^{\prime} \left( v_0 \right) \rho_0 q^2 
\tilde{V} \left( \vec{q} \right) \sin^{2}\theta & = & 0.
\label{eq:dispersion-equation-long} 
\end{eqnarray}
Let us consider the two cases of the wave vectors parallel and 
perpendicular to the 
$\hat{v}$-direction. 
For the parallel case, i.e., $\theta = 0$, 
$\sigma^{\prime} = 0$ 
or $\sigma^{\prime} = \left[ v_0 f^{\prime} \left(v_0 \right)
\pm \sqrt{\left( v_0 f^{\prime} \left(v_0 \right) \right)^{2}
- \rho_0 q^2 \tilde{V} \left( \vec{q} \right)} \right] / 2$.
On the other hand, in the 
perpendicular
case ($\theta = \pi/2$), 
$\sigma^{\prime} = v_0 f^{\prime} \left( v_0 \right)$ or 
$\sigma^{\prime} = \pm \sqrt{-\rho_0 q^2 \tilde{V} \left( \vec{q} \right)}$.
If $f^{\prime} \left( v_0 \right) > 0$, 
the homogeneous solutions are always unstable. 
For our Rayleigh-type dissipation, 
$f^{\prime} \left( v_0 \right) = -2\beta v_0 < 0$;
hence, the homogeneous solution is stable only 
when $\rho_0 q^2 \tilde{V} \left( \vec{q} \right) > 0$,
which is the same as the criterion in Eq.\,(\ref{eq:linear-stable-criterion}).
Further analysis shows that for a general angle $\theta$, 
Eq.\,(\ref{eq:dispersion-equation-long}) can be rewritten as 
$
\left( \sigma^{\prime} - v_0 f^{\prime} \left( v_0 \right) \right) 
\left( {\sigma^{\prime}}^{2} 
+ \rho_0 q^2 \tilde{V} \left( \vec{q} \right) \right)
+ \Gamma \cos^{2} \theta = 0
$, 
where $\Gamma \equiv 
v_0 f^{\prime} \left( v_0 \right) \rho_0 q^2 \tilde{V} \left( \vec{q} \right)$.
In our model, $\Gamma > 0$ whenever the homogeneous solution is unstable.
Thus, an inspection of the above equation shows that its largest root, 
i.e., the fastest growth rate, is at $\theta = \pi/2$.
As a result, perturbations on the direction perpendicular to
the swarm velocity are the 
fastest growing mode, and their rate is 
$\sqrt{-\rho_0 q^2 \tilde{V} \left( q \right)}$ for a given $q$.

\begin{figure}
\includegraphics[%
  width=0.9\columnwidth]{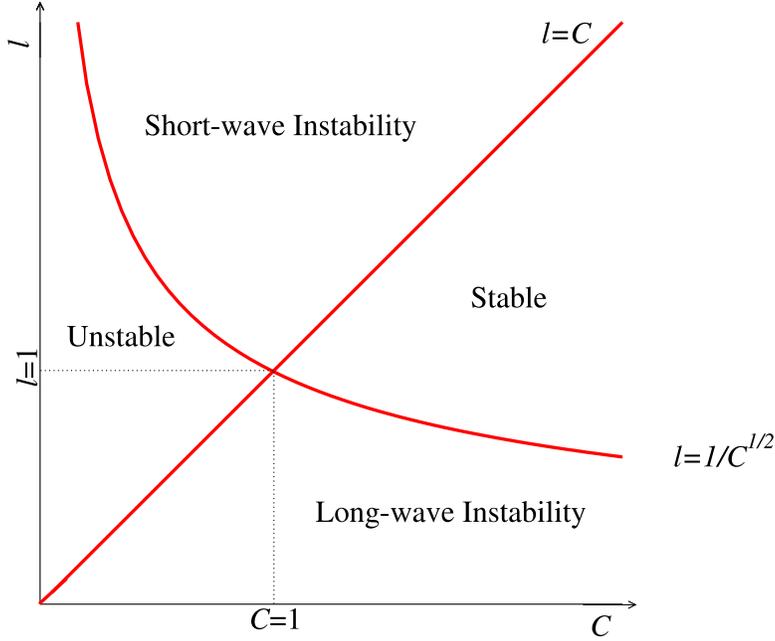}

\caption{\label{cap:linear-stability-diagram}The linear stability diagram
of the swarming model of Eqs.\,(\ref{eq:OurPosEq})\,-\,(\ref{eq:interaction_term}).
 }
\end{figure}

\begin{figure}
\includegraphics[%
  width=0.60\columnwidth]{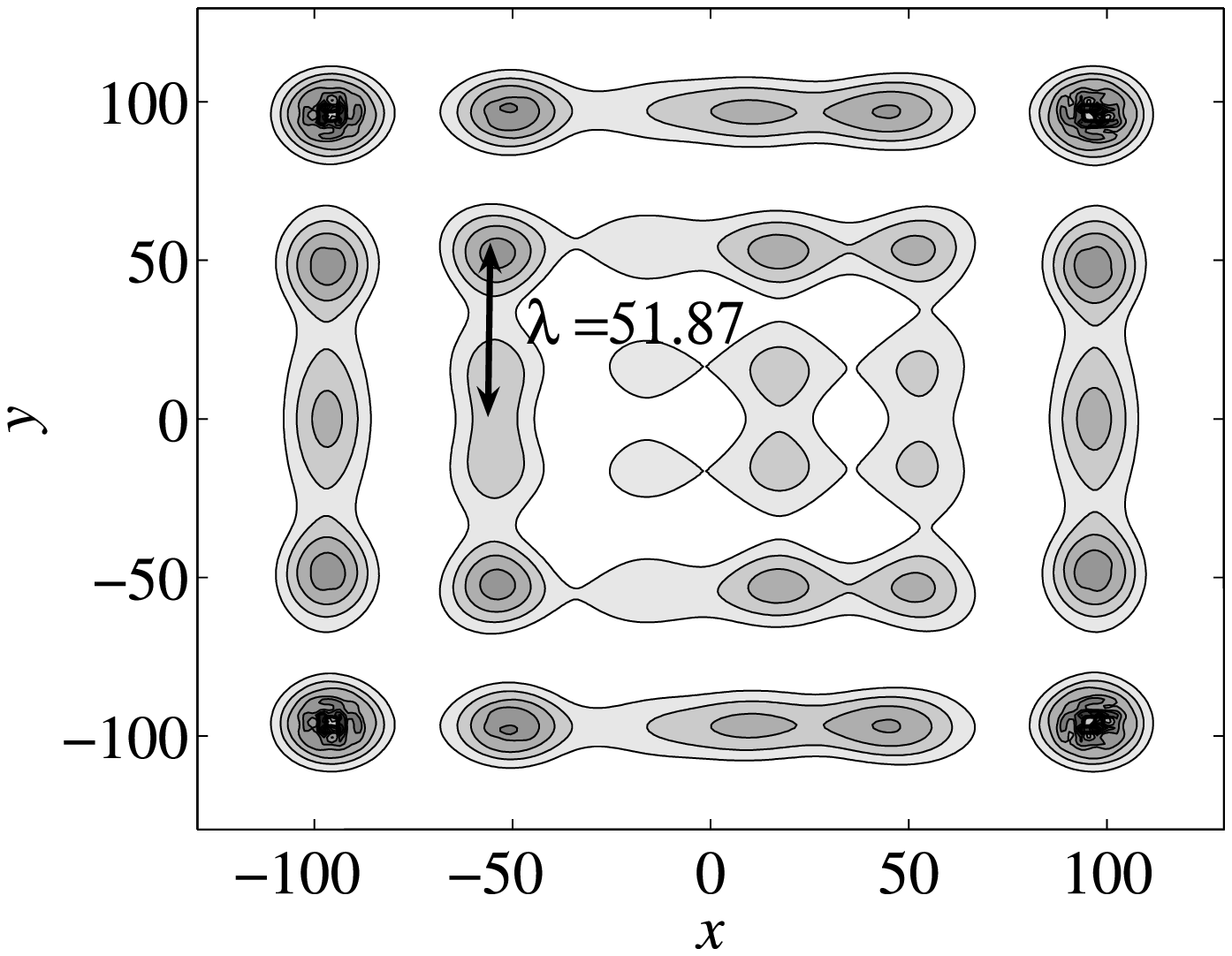}\includegraphics[%
  width=0.39\columnwidth]{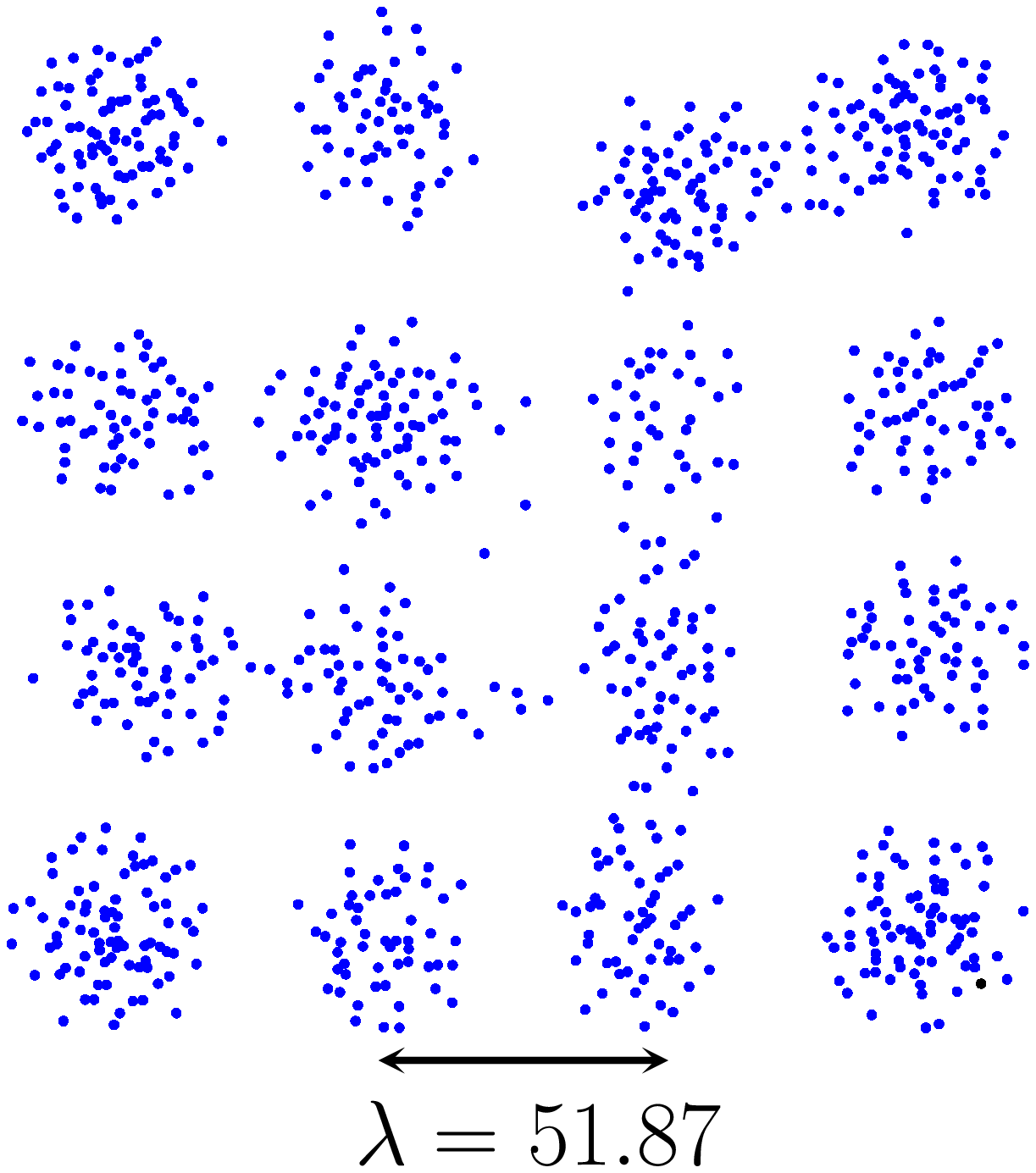}

\includegraphics[%
  width=0.60\columnwidth]{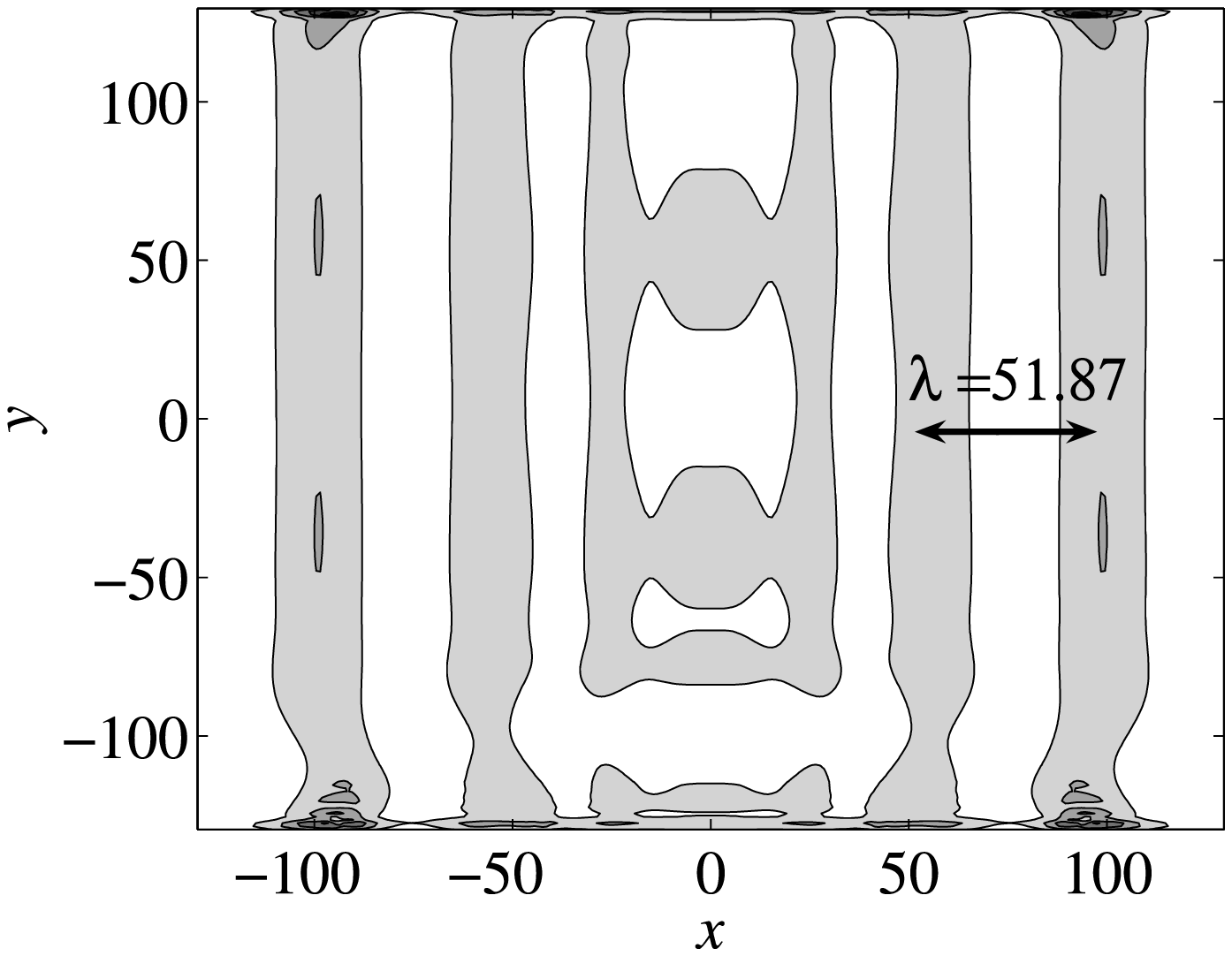}\includegraphics[%
  width=0.39\columnwidth]{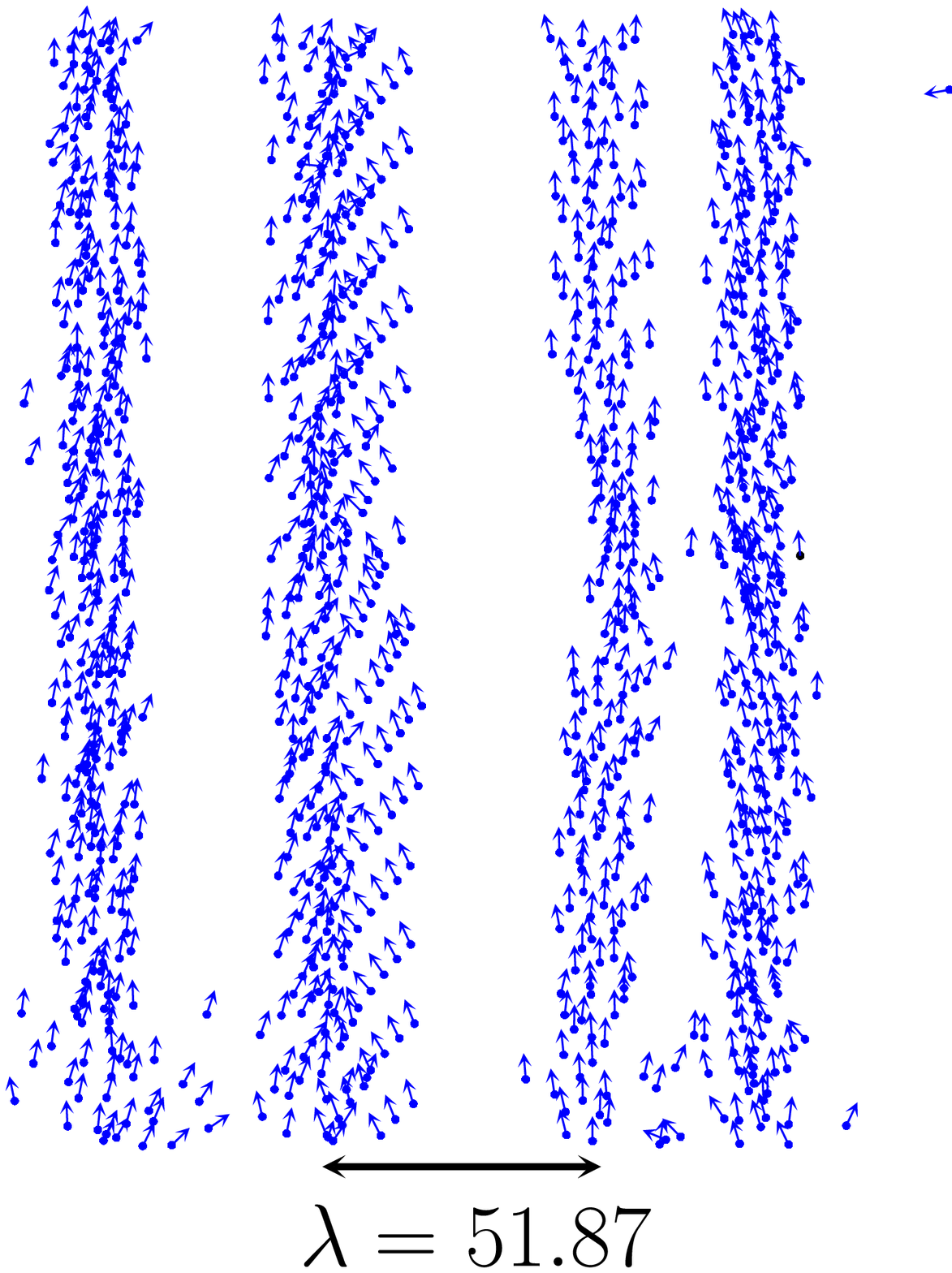}

\caption{\label{cap:onset-density-spd0}Left panel: The contours of a density
distribution of the continuum model near the instability onset with (upper) 
$\alpha=0$ and (lower) $\alpha=1$; 
Right panel: Simulations of the individual-based model using the same 
parameters and initial conditions as the left figures. 
The parameter values are $\beta=0.5$, $C_a=0.5$, $C_r=1.0$,
$\ell_a=2.0$, and $\ell_r=1.35$. 
}
\end{figure}

Substituting Eq.\,(\ref{eq:interaction_term}) into 
Eq.\,(\ref{eq:linear-stable-criterion}), the linear stability criterion 
for our swarming model can be explicitly obtained as
\begin{eqnarray}
\tilde{V} \left( q \right) & \equiv & 
2 \pi \left[ - \frac{1}{\left( 1 + {q^{\prime}}^{2} \right)^{3/2}}
+ \frac{C \ell^2}{\left( 1 + \ell^2 {q^{\prime}}^{2} \right)^{3/2}} \right] 
 >  0, 
\label{eq:linstab-criterion}
\end{eqnarray}
where $q^{\prime} \equiv q \ell_a$.
Since the above criterion has to hold for all $q^{\prime} \in \Rset$,
stability is attained at
\begin{eqnarray*}
 & \begin{array}{ccc}
   C \ell^2 > 1 &
   &
   \textrm{if $\ell < 1$}, \\
   C > \ell &
   &
   \textrm{if $\ell \ge 1$}, 
   \end{array} &  
\end{eqnarray*}
The linear stability diagram is shown in 
Fig.\,\ref{cap:linear-stability-diagram}. 
Note the close connection between the different regimes shown here and in
the H-stability diagram of Fig.\,\ref{cap:H-stab-diag}.
When the homogeneous solution is linearly stable, the interaction potential
is also H-stable.
This is because the condition of
 Eq.\,(\ref{eq:linear-stable-criterion}) is also  
sufficient, but not necessary, for H-stability \cite{Ruelle}. 
Further study on the dispersion relation in 
Eq.\,(\ref{eq:dispersion-equation-long}) reveals that 
$\sigma^{\prime}$ increases as $q^2 \tilde{V} \left( \vec{q} \right)$
decreases, and the maximum of $\sigma^{\prime}$ occurs when the minimum
of $q^2 \tilde{V} \left( \vec{q} \right)$ is reached. 
As a result, we are able to evaluate the wavelength of the fastest growth mode
and categorize the long-wave and the short-wave instability regions
in the parameter space. 
Furthermore, we compare the fastest growth wavelength to the pattern of 
the fully nonlinear continuum model near the onset of the instability,
shown in the left panel of Fig.\,\ref{cap:onset-density-spd0}. 
The simulations are initiated with a homogeneous density distribution and 
computed on a periodic domain of a $206.8 \times 206.8$ box.
The wavenumber of the fastest growth mode is calculated as 
the minimum of Eq.\,(\ref{eq:linstab-criterion}).
For the parameters chosen in Fig.\,\ref{cap:onset-density-spd0},
$\left| \vec{q} \right| = 0.121$,  
which corresponds to a wavelength $\lambda = 51.87$. 
This value matches the density aggregation patterns quite well. 
In the upper figure, $\alpha=0$; the steady state density has zero velocity,
and the $x$-$y$ directions are isotropic. 
In the lower figure, $\alpha \ne 0$, and the velocity field of the swarm 
is initiated as $\vec{u} \left( t=0 \right) = \sqrt{\alpha/\beta} \hat{y}$.
The direction of the stripes indicates that the fastest growth mode 
is indeed perpendicular to the initial velocity, which is also consistent with
the theoretical prediction.
The results can also be compared to the simulation of the individual-based
model by using the same parameter values and equivalent initial and boundary
conditions.
Since $V \left( r \right)$ decays rapidly in $r$, $U \left( \vec{x}_i \right)$
can be well approximated by including only the adjacent eight boxes 
surrounding the computational domain.
The steady particle distributions of the individual-based simulations are shown
on the right panel of Fig.\,\ref{cap:onset-density-spd0}.
The theoretically predicted wavelength agrees with the patterns seen  
in the simulations of the continuum and the individual-based models. 

\section{Discussion \label{Sec-Discussion}}
Soft-core interactions are widely adopted in the swarming literature 
\cite{TonerTuSwarming,Brownian1,LevineModel,CouzinDiscreteModel,Brownian2}.
Our investigations, with the Morse potential of Eq.\,(\ref{eq:interaction_term}),
reveal that the commonly observed 
core-free
mill patterns only exist in the catastrophic regime and not in the H-stable
one. 
In this latter case, particles arrange in rigid-body-like structures, 
similar to other H-stable interactions such as the Lennard-Jones 
potential.
The morphology richness associated with soft-core catastrophic potentials
is one of the reasons that such interactions have been so broadly applied
in the literature.
However, a soft-core interaction, such as Eq.\,(\ref{eq:interaction_term}), 
does not prevent particles from occupying the same space, 
which is an unphysical situation. 
This can be resolved in two ways. 
One is that animals usually flock on a reduced dimension and thus, 
can use the extra dimension to avoid actually occupying the same space. 
For example, ants can crawl over each other; therefore, they can use 
$z$-direction to ``pass through'' each other when they flock on the
$x$-$y$ plane. 
Another way is to actually add an additional hard-core repulsion solely 
to prevent overlapping. 
In other words, there is a soft-core potential that defines an equilibrium 
distance between particles and gives rise to the swarming patterns, 
and there is also a hard-core potential that specifies a forbidden distance 
and prevent particles from penetrating each other.
We find that the presence of the hard-core potential affects the swarming 
pattern only when $N$ is large enough, and hence the equilibrium distance 
between nearest neighbors, determined by the collective soft-core interaction,
collapses to the vicinity of the hard-core forbidden distance. 
Otherwise, flocks exhibit the same soft-core steady state patterns for 
small to moderate $N$, 
except for the double-mill state, 
which is apparently very sensitive to hard-cores and, in our simulations, are 
absent altogether.
As $N$ increases, the equilibrium $\delta_\mathrm{NND}$ at first decreases;
the flock size increases with $N$ only when the equilibrium 
$\delta_\mathrm{NND}$ becomes close to the forbidden hard-core zone and 
cannot decrease further. 
Thus, a swarming flock at moderate $N$ can have soft-core patterns 
in spite of the existence of a local hard-core repulsion.

Despite the natural tendency to keep a reasonable distance between 
each other, animals may still come close and occasionally touch 
each other while moving in a biological swarm. 
Thus, the natural repulsive tendency can be realized as a soft-core 
repulsion while the body length of the swarming animals can be viewed 
as a hard-core forbidden zone. 
For biological swarms, the equilibrium $\delta_\mathrm{NND}$ is visibly larger
than the hard-core forbidden zone, which supports the description given
in the previous paragraph.
In contrast, the Lennard-Jones potential, used for physical systems of
molecules, defines an equilibrium distance very close to where the potential 
rapidly rises toward infinity. 
In other words, the equilibrium distance is nearly the same as the hard-core
forbidden zone.
Compressibility is perhaps the reason why various catastrophic patterns, 
which are not observed in the condensed phases of classical matter, 
can exist in the aggregation states of natural swarms.
In artificial swarms, the hard-core repulsion can be understood as a 
collision avoidance strategy. 
If the distance to invoke the collision avoidance is 
much shorter than the equilibrium spacing between agents,
various collapsing patterns shown in Ref.\,\cite{dorsogna} become possible
and might even be engineered for artificial swarming of vehicles.

\section{Summary \label{sec:summary}}
Natural swarms may switch patterns under different circumstances.
While the changes of pattern have been understood as a result of changing
individual mobility and mutual interactions within the swarm, 
our individual-based model similarly exhibits a state transition 
through various swarming patterns by varying the parameters.
The same idea can be applied to artificial swarms, where a group of robots 
can be programmed to strategically change formations by varying 
the self-driving and communicating parameters of the control model.
To analyze the stability of the emerging patterns with respect to 
the model parameters, it is advantageous to have a continuum model 
that can precisely describe the individual-based model. 
We illustrate a procedure to derive a continuum model from an 
individual-based model by using classical statistical mechanics.
We show that the derived continuum model does not approximate
the individual dynamics when the interaction potential is H-stable.
This is due to the fact that for H-stable systems, the length scale of the 
potential is comparable to interparticle distances,
whereas in the catastrophic regime
many particles can co-exist on a length scale comparable to the scale
of the potential.
In the catastrophic regime, 
the steady state solution of the continuum model well matches the 
single-mill pattern of the individual-based model.
The long-wave instability also shows a match to both the continuum 
and the individual-based model simulations when we theoretically 
analyze the linear stability of the homogeneous solution for the 
continuum model.
Thus, the continuum model may be useful for further analysis, 
such as the stability of non-trivial solutions.

\vspace{5mm}\noindent{\bf Acknowledgment}.  
{\it The authors thank Herbert Levine, Jianhong Shen, and Chad M. Topaz 
for useful discussions.
We acknowledge support from ARO grant W911NF-05-1-0112, 
ONR grant N000140610059, 
and NSF grant DMS-0306167.}
\vspace{5mm}

\appendix
\section{Derivation of the fluctuation terms}
Following Ref. \cite{IrvingKirkwood},
the momentum transport equation can be obtained by substituting the
macroscopic momentum 
\[
\rho \left( \vec{x},t \right) \vec{u} \left( \vec{x},t \right)
 = \left\langle \sum_{i=1}^{N}
{\vec{p}_i \delta \left( \vec{x}_i - \vec{x} \right)};f \right\rangle 
\]
into the generalized Liouville Equation, 
valid for non-conserved systems \cite{Tuckerman},
\begin{eqnarray*}
\frac{\partial \left( \rho \vec{u} \right)}{\partial t} & = & 
\frac{\partial}{\partial t} \left\langle \sum_{i=1}^{N}
{\vec{p}_i \delta \left( \vec{x}_i - \vec{x} \right)};f \right\rangle \\
 & = & \sum_{k=1}^{N}
{\left\langle \frac{\vec{p}_k}{m} \cdot \vec{\nabla}_{\vec{x}_k} \left(
\sum_{i=1}^{N}{\vec{p}_i \delta \left(\vec{x}_i - \vec{x} \right)}
\right) 
+ \dot{\vec{p}}_k \cdot \vec{\nabla}_{\vec{p}_k} \left(
\sum_{i=1}^{N}{\vec{p}_i \delta \left( \vec{x}_i - \vec{x} \right)} 
\right);f \right\rangle }.
\end{eqnarray*}
Here $f$ is the probability density function described in 
Eq.\,(\ref{eq:canonical-pdf}).
Since 
\begin{eqnarray*}
\frac{\vec{p}_k}{m} \cdot \vec{\nabla}_{\vec{x}_k} \left(
\sum_{i=1}^{N}{\vec{p}_i \delta \left( \vec{x}_i - \vec{x} \right)}
\right) & = & \frac{\vec{p}_k}{m} \cdot \vec{\nabla}_{\vec{x}_k} \vec{p}_k 
\delta \left( \vec{x}_k - \vec{x} \right) \\
 & = & -\vec{\nabla}_{\vec{x}} 
\cdot \left( \frac{\vec{p}_k \vec{p}_k}{m} \right)
\delta \left( \vec{x}_k - \vec{x} \right), \\
\dot{\vec{p}}_k \cdot \vec{\nabla}_{\vec{p}_k} \left( 
\sum_{i=1}^{N}{\vec{p}_i \delta \left( \vec{x}_i - \vec{x} \right)}
\right) & = & \dot{\vec{p}}_k \delta \left( \vec{x}_k - \vec{x} \right),
\end{eqnarray*}
the transport equation can further be reduced to
\begin{eqnarray}
\frac{\partial \left( \rho \vec{u} \right)}{\partial t} & = & \sum_{k=1}^{N}
{\left[ -\nabla_{\vec{x}} \cdot \left\langle 
\left( \frac{\vec{p}_k \vec{p}_k}{m} \right) 
\delta \left( \vec{x}_k - \vec{x} \right);f 
\right\rangle  
 + \left\langle 
\dot{\vec{p}}_k \delta \left(\vec{x}_k - \vec{x} \right);f 
\right\rangle \right] }. 
\label{eq:transport-temp1}
\end{eqnarray}
The first term on the right hand side can be modified by noting that
\begin{eqnarray*}
 &  & \sum_{k=1}^{N}{m \left\langle 
\left( \frac{\vec{p}_k}{m} - \vec{u} \right) 
\left( \frac{\vec{p}_k}{m} - \vec{u} \right)
\delta \left( \vec{x}_k - \vec{x} \right);f \right\rangle } \\
= &  & \sum_{k=1}^{N}{\left\langle 
\left( \frac{\vec{p}_k \vec{p}_k}{m} \right) 
\delta \left(\vec{x}_k - \vec{x} \right);f \right\rangle } 
 - \vec{u} \sum_{k=1}^{N}{\left\langle 
\vec{p}_k \delta \left( \vec{x}_k - \vec{x} \right);f \right\rangle } \\
 & - & \sum_{k=1}^{N}{\left\langle 
\vec{p}_k \delta \left( \vec{x}_k - \vec{x} \right);f \right\rangle } \vec{u}
 + \vec{u} \vec{u} \sum_{k=1}^{N}
{\left\langle m \delta \left(\vec{x}_k - \vec{x} \right);f \right\rangle } \\
= &  & \sum_{k=1}^{N}{\left\langle \left(\frac{\vec{p}_k \vec{p}_k}{m} \right)
\delta \left( \vec{x}_k - \vec{x} \right);f \right\rangle } 
 - \rho \vec{u} \vec{u},
\end{eqnarray*}
where $\vec{u}$ is the macroscopic velocity defined in 
Eq.\,(\ref{eq:continuum velocity}).
Eq.\,(\ref{eq:transport-temp1}) then becomes
\begin{eqnarray}
\frac{\partial \left( \rho \vec{u} \right)}{\partial t} 
+ \vec{\nabla}_{\vec{x}} \cdot \left( \rho \vec{u} \vec{u} \right) & = & 
- \vec{\nabla}_{\vec{x}} \cdot \hat{\sigma}_\mathrm{K} \left( \vec{x},t \right)
+\sum_{k=1}^{N}{\left\langle 
\dot{\vec{p}}_k \delta \left( \vec{x}_k - \vec{x} \right);f 
\right\rangle },
\label{eq:transport-temp2} 
\end{eqnarray}
where 
\begin{eqnarray*}
\hat{\sigma}_\mathrm{K} & = & \sum_{k=1}^{N}{m \left \langle 
\left( \frac{\vec{p}_k}{m} - \vec{u} \right) 
\left( \frac{\vec{p}_k}{m} - \vec{u} \right)
\delta \left( \vec{x}_k - \vec{x} \right);f \right\rangle }.
\end{eqnarray*}
We can substitute the explicit form of $\dot{\vec{p}}_k$ from 
Eqs.\,(\ref{eq:continuum velocity}) and (\ref{eq:continuum-force})
\begin{eqnarray*}
\dot{\vec{p}}_k & = &
\alpha\vec{p}_k - 
\beta \frac{\left| \vec{p}_k \right|^{2}}{m^2} \vec{p}_k
 - \vec{\nabla} U \left( \vec{x}_k \right)
\end{eqnarray*}
into the second term of Eq.\,(\ref{eq:transport-temp2}) 
\begin{eqnarray*}
\sum_{k=1}^{N}{\left\langle \dot{\vec{p}}_k 
\delta \left( \vec{x}_k - \vec{x} \right);f \right\rangle } & = & 
\sum_{k=1}^{N}{\left\langle \left( \alpha\vec{p}_k - 
\beta \frac{\left| \vec{p}_k \right|^{2}}{m^2} \vec{p}_k
 - \vec{\nabla} U \left( \vec{x}_k \right) \right)
\delta \left( \vec{x}_k - \vec{x} \right);f \right\rangle } \\
 & = & \alpha \rho \vec{u} - \sum_{k=1}^{N}{\left\langle \left( \beta 
\frac{\left| \vec{p}_k \right|^2}{m^2} \vec{p}_k \right) \right\rangle} 
 + \vec{F}_V.
\end{eqnarray*}
The second term above can be further simplified as
\begin{eqnarray*}
 &  & \sum_{k=1}^{N}{\left\langle 
\left( \beta \frac{\left| \vec{p}_k \right|^2}{m^2} \vec{p}_k \right) 
\delta \left( \vec{x}_k - \vec{x} \right);f \right\rangle } = \\
 &  & \beta \sum_{k=1}^{N}{\left\langle 
\left( \frac{\left| \vec{p}_k \right|^2}{m^2} \vec{p}_k \right) 
\delta \left(\vec{x}_k - \vec{x} \right);f \right\rangle } 
 -  \beta \sum_{k=1}^{N}{\left\langle 
\frac{\left| \vec{p}_k \right|^2}{m} \vec{u} 
\delta \left( \vec{x}_k - \vec{x} \right);f \right\rangle } \\
 & + & \beta \sum_{k=1}^{N}{\left\langle m \left( -2 \frac{\vec{p}_k}{m} 
\cdot \vec{u} + \left| \vec{u} \right|^2 \right)  
\left( \frac{ \vec{p}_k}{m} - \vec{u} \right)
\delta \left( \vec{x}_k - \vec{x} \right);f \right\rangle } \\
 & + & 2\beta E_\mathrm{K} \vec{u} 
 - 2\beta \vec{u} \cdot \hat{\sigma}_\mathrm{K} 
 + \beta \sum_{k=1}^{N}{\left \langle m \left| \vec{u} \right|^2 
\left( \frac{\vec{p}_k}{m} - \vec{u} \right) 
\delta \left( \vec{x}_k - \vec{x} \right);f \right\rangle } \\
= &  & 2 \beta \sum_{k=1}^{N}{\left\langle \frac{m}{2} 
\left| \frac{\vec{p}_k}{m} - \vec{u} \right|^{2} 
\left( \frac{\vec{p}_k}{m} - \vec{u} \right) 
\delta \left( \vec{x}_k - \vec{x} \right);f \right\rangle } 
 + 2 \beta E_\mathrm{K} \vec{u} 
 - 2 \beta \vec{u} \cdot \hat{\sigma}_\mathrm{K} \\
= &  & 2 \beta \vec{q}_\mathrm{K} + 2 \beta E_\mathrm{K} \vec{u}
- 2 \beta \vec{u} \cdot \hat{\sigma}_\mathrm{K},
\end{eqnarray*}
where 
\[
\vec{q}_\mathrm{K} = \sum_{i=1}^{N}{\left\langle 
\frac{m}{2} \left| \frac{\vec{p}_i}{m} - \vec{u} \right|^2 
\left( \frac{\vec{p}_i}{m} - \vec{u} \right)
\delta \left( \vec{x}_i - \vec{x} \right);f \right\rangle }.
\]
As a result, Eq.\,(\ref{eq:transport-temp2}) can be written as 
\begin{eqnarray*}
\frac{\partial}{\partial t} \left( \rho \vec{u} \right)
+ \vec{\nabla} \cdot \left( \rho \vec{u} \vec{u} \right) 
+ \vec{\nabla} \cdot \hat{\sigma}_\mathrm{K} 
& = & 
\alpha \rho \vec{u} - 2 \beta E_\mathrm{K} \vec{u} 
-2 \beta \vec{q}_\mathrm{K} + 2 \beta \vec{u} \cdot \hat{\sigma}_\mathrm{K} 
+ \vec{F}_V,
\end{eqnarray*}
which is the momentum transport equation shown in Eq.\,(\ref{eq:mom-tran-eqn0}).


\bibliographystyle{elsart-num}
\bibliography{yaoli.bib}

\begin{thebibliography}{10}
\expandafter\ifx\csname url\endcsname\relax
  \def\url#1{\texttt{#1}}\fi
\expandafter\ifx\csname urlprefix\endcsname\relax\def\urlprefix{URL }\fi

\bibitem{Vicsek-phase}
T.~Vicsek, A.~Czirk, E.~Ben-Jacob, I.~Cohen, O.~Shochet, Novel type of phase
  transition in a system of self-driven particles, Phys. Rev. Lett. 75 (1995)
  1226--1229.

\bibitem{TonerTuSwarming}
J.~Toner, Y.~Tu, Long-range order in a two-dimensional dynamical xy model: how
  birds fly together, Phys. Rev. Lett. 75 (1995) 4326--4329.

\bibitem{Mogilner96}
A.~Mogilner, L.~Edelstein-Keshet, Spatio-angular order in populations of
  self-aligning objects: formation of oriented patches, Physica D 89 (1996)
  346--367.

\bibitem{Sugawara97}
K.~Sugawara, M.~Sano, Cooperative acceleration of task performance: Foraging
  behavior of interacting multi-robots system, Physica D 100 (1997) 343--354.

\bibitem{ScienceSwarming}
J.~Parrish, L.~Edelstein-Keshet, Complexity, pattern, and evolutionary
  trade-offs in animal aggregation, Science 294 (1999) 99--101.

\bibitem{NaomiUAV}
N.~E. Leonard, E.~Fiorelli, Virtual leaders, artificial potentials and
  coordinated control of groups, in: Proc. 40th IEEE Conf. Decision Contr.,
  2001, pp. 2968--2973.

\bibitem{FishSchoolSwarming}
J.~Parrish, S.~V. Viscido, D.~Gr\"unbaum, Self-organized fish schools: an
  examination of emergent properties, Biol. Bullet. 202 (2002) 296--305.

\bibitem{UAV1}
Y.~Liu, K.~M. Passino, M.~M. Polycarpou, Stability analysis of m-dimensional
  asynchronous swarms with a fixed communication topology, in: IEEE Trans.
  Autom. Contr., Vol.~48, 2003, pp. 76--95.

\bibitem{UAV2}
A.~Jadbabaie, J.~Lin, A.~S. Morse, Coordination of groups of mobile agents
  using nearest neighbor rules, in: IEEE Trans. Autom. Contr., Vol.~48, 2003,
  pp. 988--1001.

\bibitem{Niwa94}
H.~S. Niwa, Self-organizing dynamic model of fish schooling, J. Theor. Biol.
  171 (1994) 123--136.

\bibitem{Shimoyama}
N.~Shimoyama, K.~Sugawara, T.~Mizuguchi, Y.~Hayakawa, M.~Sano, Collective
  motion in a system of motile elements, Phys. Rev. Lett. 76 (1996) 3870--3873.

\bibitem{Romey1}
W.~L. Romey, Individual differences make a difference in the trajectories of
  simulated schools of fish, Ecol. Model. 92 (1996) 65--77.

\bibitem{Brownian1}
A.~S. Mikhailov, D.~H. Zanette, Noise-induced breakdown of coherent collective
  motion in swarms, Phys. Rev. E 60 (1999) 4571--4575.

\bibitem{FlierlDiscreteModel}
G.~Flierl, D.~Gr\"unbaum, S.~Levin, D.~Olson, From individuals to aggregations:
  the interplay between behavior and physics, J. Theor. Biol. 196 (1999)
  397--454.

\bibitem{LevineModel}
H.~Levine, W.~J. Rappel, I.~Cohen, Self-organization in systems of
  self-propelled particles, Phys. Rev. E 63 (2000) 017101.

\bibitem{CouzinDiscreteModel}
I.~D. Couzin, N.~R. Franks, Self-organized lane formation and optimized traffic
  flow in army ants,, Proc. R. Soc. Lond. B 270 (2002) 139--146.

\bibitem{Brownian2}
U.~Erdmann, W.~Ebeling, V.~S. Anishchenko, Excitation of rotational modes in
  two-dimensional systems of driven brownian particles, Phys. Rev. E 65 (2002)
  061106.

\bibitem{Brownian3}
W.~Ebeling, U.~Erdmann, Nonequilibrium statistical mechanics of swarms of
  driven particles, Complexity 8 (2003) 23--30.

\bibitem{Viscido1}
S.~V. Viscido, J.~K. Parrish, D.~Gr\"unbaum, Individual behavior and emergent
  properties of fish schools: a comparison of observation and theory, Mar.
  Ecol. Prog. Ser. 273 (2004) 239--249.

\bibitem{Viscido2}
S.~V. Viscido, J.~K. Parrish, D.~Gr\"unbaum, The effect of population size and
  number of influential neighbors on the emergent properties of fish schools,
  Ecol. Model. 183 (2005) 347--363.

\bibitem{Brownian4}
U.~Erdmann, W.~Ebeling, A.~Mikhailov, Noise-induced transition from
  translational to rotational motion of swarms, Phys. Rev. E 71 (2005) 051904.

\bibitem{BandSolutionSwarming}
L.~Edelstein-Keshet, J.~Watmough, D.~Gr\"unbaum, Do travelling band solutions
  describe cohesive swarms? an investigation for migratory locusts, J. Math.
  Biol. 36 (1998) 515--549.

\bibitem{Mogilner1}
A.~Mogilner, L.~Edelstein-Keshet, A non-local model for a swarm, J. Math. Biol.
  38 (1999) 534--549.

\bibitem{Oien}
A.~H. {\O}ien, Daphnicle dynamics based on kinetic theory: an
  analogue-modelling of swarming and behaviour of daphnia, Bullet. Math. Biol.
  66 (2004) 1--46.

\bibitem{TopazSwarming}
C.~M. Topaz, A.~L. Bertozzi, Swarming patterns in a two-dimensional kinematic
  model for biological groups, SIAM J. Appl. Math. 65 (2004) 152--174.

\bibitem{Topaz2}
C.~M. Topaz, A.~L. Bertozzi, M.~A. Lewis, A nonlocal continuum model for
  biological aggregation, Bullet. Math. Biol., to appear.

\bibitem{IrvingKirkwood}
J.~H. Irving, J.~G. Kirkwood, The statistical mechanical theory of transport
  processes. iv. the equations of hydrodynamics, J. Chem. Phys. 18 (1950)
  817--829.

\bibitem{dorsogna}
M.~R. D'Orsogna, Y.~L. Chuang, A.~L. Bertozzi, L.~S. Chayes, Self-propelled
  particles with soft-core interactions: patterns, stability, and collapse,
  Phys. Rev. Lett. 96 (2006) 104302.

\bibitem{Ruelle}
D.~Ruelle, Statistical Mechanics, Rigorous Results, New York: W.A. Benjamin,
  1969.

\bibitem{Schneirla}
T.~C. Schneirla, Army Ants: A Study in Social Organization, W. H. Freeman,
  1971.

\bibitem{Koch-White-myxobacteria}
A.~L. Koch, D.~White, The social lifestyle of myxobacteria, Bioessays 20 (1998)
  1030--1038.

\bibitem{Rayleigh}
J.~W.~S. Rayleigh, The Theory of Sound, 2nd Edition, Vol.~1, MacMillan, London,
  1894.

\bibitem{Weihs73}
D.~Weihs, Optimal fish cruising speed, Nature 245 (1973) 48--50.

\bibitem{Lambert:num-ode}
J.~D. Lambert, Numerical Methods for Ordinary Differential Equations, John
  Wiley \& Sons, 1991.

\bibitem{passino1}
V.~Gazi, K.~Passino, Stability analysis of swarms, in: IEEE Trans. Autom.
  Contr., Vol.~48, 2003, pp. 692--697.

\bibitem{passino2}
V.~Gazi, K.~Passino, A class of attractions/repulsion functions for stable
  swarm aggregations, in: Proc. Conf. Decision Contr., 2002, pp. 2842--2847.

\bibitem{passino3}
V.~Gazi, K.~Passino, Stability analysis of social foraging swarms: combined
  effects of attractant/repellent profiles, in: Proc. Conf. Decision Contr.,
  2002, pp. 2848--2853.

\bibitem{granular}
C.~S. Campbell, Rapid granular flows, Annu. Rev. Fluid Mech. 22 (1990) 57--92.

\bibitem{Grunbaum1}
D.~Gr\"unbaum, Translating stochastic density-dependent individual behavior
  with sensory constraints to an eulerian model of animal swarming, J. Math.
  Biol. 33 (1994) 139--161.

\bibitem{Mourelo1}
P.~G\'omez-Mourelo, From individual-based models to partial differential
  equations. an application to the upstream movement of elvers, Ecol. Model.
  188 (2005) 93--111.

\bibitem{Morale1}
D.~Morale, V.~Capasso, K.~Oelschl\"ager, An interacting particle system
  modelling aggregation behavior: from individuals to populations, J. Math.
  Biol. 50 (2005) 49--66.

\bibitem{Tuckerman}
M.~E. Tuckerman, C.~J. Mundy, M.~L. Klein, Toward a statistical thermodynamics
  of steady states, Phys. Rev. Lett. 78 (1997) 2042--2045.

\bibitem{Leveque:num-pde}
R.~J. Leveque, Numerical Methods for Conservation Laws, 1st Edition,
  Birkh\"auser, 1992.

\end{thebibliography}


\end{document}